\documentclass{aastex63}

\usepackage{amsmath,amssymb,bm}
\usepackage{lineno}
\usepackage{color}
\definecolor{ForrestGreen}{rgb}{0.133,0.545,0.133}
\definecolor{DarkGreen}{rgb}{0.0,0.45,0.0}
\newcommand{\B}[1]{{\color{blue}{\textbf{#1}}}} \newcommand{\R}[1]{{\color{red}{\textbf{#1}}}} \newcommand{\G}[1]{{\color{DarkGreen}{\textbf{#1}}}} \newcommand{\M}[1]{{\color{magenta}{\textbf{#1}}}}

\begin{document}

\title{Formation and Eruption of Hot Channels during an M6.5 Class Solar Flare}

\correspondingauthor{Yingna Su}
\email{ynsu@pmo.ac.cn}

\author{Yanjie Liu}
\affiliation{Key Laboratory of Dark Matter and Space Astronomy, Purple Mountain Observatory, Chinese Academy of
	Sciences, Nanjing, Jiangsu 210023, China}

\affiliation{Department of Astronomy and Space Science, University of Science and Technology of China, Hefei, Anhui 230026, China} 

\author{Yingna Su}
\affiliation{Key Laboratory of Dark Matter and Space Astronomy, Purple Mountain Observatory, Chinese Academy of
	Sciences, Nanjing, Jiangsu 210023, China}

\affiliation{Department of Astronomy and Space Science, University of Science and Technology of China, Hefei, Anhui 230026, China} 

\author{Rui Liu}
\affiliation{CAS Key Laboratory of Geospace Environment, Department of Geophysics and Planetary Sciences, University of Science and Technology of China, Hefei 230026, China}

\author{Jialin Chen}
\affiliation{Key Laboratory of Dark Matter and Space Astronomy, Purple Mountain Observatory, Chinese Academy of Sciences, Nanjing, Jiangsu 210023, China}

\affiliation{Department of Astronomy and Space Science, University of Science and Technology of China, Hefei, Anhui 230026, China}

\author{Tie Liu}
\affiliation{School of Astronomy and Space Science, Nanjing University, Nanjing, Jiangsu 210033, China}

\affiliation{Key Laboratory for Modern Astronomy and Astrophysics (Nanjing University), Ministry of Education, Nanjing, Jiangsu 210033, China}

\author{Haisheng Ji}
\affiliation{Key Laboratory of Dark Matter and Space Astronomy, Purple Mountain Observatory, Chinese Academy of
	Sciences, Nanjing, Jiangsu 210023, China}

\affiliation{Department of Astronomy and Space Science, University of Science and Technology of China, Hefei, Anhui 230026, China}



\begin{abstract}
We investigate the formation and eruption of hot channels associated with the M6.5 class
flare (SOL2015-06-22T18:23) occurring in NOAA AR 12371 on 2015 June 22. Two flare 
precursors are observed before the flare main phase. Observations in 94~\AA~and 
131~\AA~by SDO/AIA have revealed the early morphology of the first hot channel as a 
group of hot loops, which is termed as seed hot channel. A few seed hot channels are 
formed above the polarity inversion line (PIL) and the formation is associated with 
footpoint brightenings' parallel motion along the PIL, which proceeds into the early 
stage of the flare main phase. During this process, seed hot channels build up and rise 
slowly, being accelerated at the peak of the second precursor. They merge in the process 
of acceleration forming a larger hot channel, which then forms an “inverted $\gamma$” 
shape kinking structure. Before the flare peak, the second kinking hot channel with 
negative crossing appears near the first kinking hot channel that has erupted. The 
eruption of these two hot channels produce two peaks on the main flare's GOES light 
curve. The footpoint brightenings' propagation along the PIL indicate that the first 
kinking hot channel may be formed due to zipper reconnection. The occurrence of merging 
between seed hot channels observed by AIA is supported by the extrapolated nonlinear 
force-free field models. The observed writhing motion of the first kinking hot 
channel may be driven by the Lorentz force.
\end{abstract}

\keywords{Sun: corona --- Sun: flare --- Sun: filament --- Sun: magnetic field}

\section{Introduction}

Solar flares, prominence/filament eruptions and coronal mass ejections (CME) are the 
three most intense solar activities. The high-speed magnetized plasma contained in  
magnetic flux ropes (MFRs) released during these solar activities may interact with the 
magnetosphere and ionosphere, and seriously affect the safety of human high-tech 
infrastructures. The MFR, a set of helical magnetic field lines wrapping around a 
common axis, is generally considered to be the fundamental structure in the CME/flare 
dynamical process \citep{Shibata1995,Zhang2012,Cheng2016,Liu2020}. 

The formation of MFR is still an ongoing debate \citep{Patsourakos2020}. Some 
studies believe that the MFR is present before the eruption
\citep{Kopp1976,Forbes1995,Titov1999}. This pre-existent MFR may be formed in the 
convection zone, but it is forced by magnetic buoyancy to emerge into the corona 
\citep{Rust1994,Fan2003,Fan2009,Aulanier2012}. Or it is formed due to the slow magnetic 
reconnection between sheared arcades in the low corona 
\citep{Green2009,Liu2010,Green2011,Patsourakos2013}, that is, the shear motion in the 
photosphere leads to the evolution of the potential field into sheared magnetic field, 
and as the magnetic field shear increases, the sheared arcades reconnect and the MFR is 
formed after multiple reconnections \citep{vanBallegooijen1989}. Other studies suggest 
that the MFR is formed during the eruption process. In this case, only sheared arcades 
exist before eruption. During the eruption, an MFR is formed due to magnetic 
reconnection between sheared arcades, which then continuously increases the magnetic 
flux of the MFR \citep{Antiochos1999,Moore2001,Karpen2012}. Recent studies 
demonstrate that there may exist a ``hybrid" scenario, in which a seed MFR forms within 
the sheared arcades via tether-cutting reconnection before eruption, and builds up 
via flare reconnection during the eruption \citep{Gou2019,Liu2020}. 
\citet{Patsourakos2020} also suggest the pre-eruptive configuration as a 
``hybrid" state consisting of both sheared arcades and MFRs, with different 
configurations at different evolutionary stages.

According to whether the magnetic reconnection is involved or not, the triggering 
mechanism of MFR eruption can be divided into two types. One is 
magnetohydrodynamic (MHD) instability, including torus instability (TI, 
\citealt{Kliem&Torok2006}) and kink instability (KI, 
\citealt{Fan2003,Fan2004,Torok2004}). In addition, the double arc instability 
(DAI) is considered to be able to drive the early stage of the eruption
\citep{Ishiguro2017,Kusano2020}. The other type is based on magnetic reconnection. 
Magnetic reconnections occurring above and below the MFR, correspond to the break-out 
model \citep{Antiochos1999,Lynch2008} and the tether-cutting model 
\citep{Moore2001,Jiang2021}, respectively. The formation time of MFR is also different 
in these mechanisms. In the MHD model, the MFR already exists before the
eruption. In the tether-cutting model, the MFR is formed 
prior to the slow rise phase and grows up in the acceleration phase 
\citep{Moore2001}. While the MFR is not formed until the acceleration phase in 
the break-out model \citep{Karpen2012,Cheng2016}.

There are many observational structures in the solar atmosphere related to MFRs, 
including the sigmoids \citep{Green&Kliem2014}, filaments and filament channels 
\citep{vanBallegooijen1998,Mackay2010,Su2012}, hot channels 
\citep{Cheng2011,Zhang2012} and coronal cavities \citep{Low1995}. As a new  
evidence of MFR observation, hot channels usually appear as EUV blobs in the high 
temperature bands and as dark cavities in the low temperature bands 
\citep{Cheng2016,Cheng2017}. The height evolution of the hot channel can be fitted with 
a function consisting of linear and exponential components, which corresponds to the 
slow rise and the impulsive acceleration of the hot channel, respectively. 

The hot channel shows a remarkable morphological evolution in the early stage of the 
eruption. It often appears as a twisted and/or writhed sigmoidal structure, and then 
transforms into a semi-circular shape in the slow rise phase, after which the impulsive 
accelerate begins \citep{Zhang2012}. Writhe is related to the rotation of erupting MFR, 
and the mechanism governing this rotation has received extensive attention. When the KI 
occurs, the axis of the MFR rotates rapidly, transforming part of the twist into writhe 
\citep{Baty2001,Torok2014}. While recent MHD simulations find that writhe can 
be transformed into twist along with the rotation of MFR \citep{Zhou2022}. In 
addition, the external sheared field can also contribute to the rotation of the MFR 
\citep{Isenberg2007,Lynch2009}. And the rotations caused by these two mechanisms point 
in the same direction. The rotation by the external sheared field tends to distribute 
across a larger height range, and if the sources of the external stabilizing field have 
a smaller distance than the footpoints of the erupting flux, the external sheared field 
yields the major contribution to the rotation. The rotation due to twist relaxation 
tends to work mainly in the low corona with a height range up to several times of the 
distance between the footpoint of the MFR \citep{Kliem2012}. In addition, magnetic 
reconnection with the surrounding magnetic field \citep{Shiota2010}, the straightening 
from the initial S-shape \citep{Torok2010}, and the asymmetric deflection of the rising 
flux during propagation through the overlying field can all contribute to the rotation 
of the MFR \citep{Yurchyshyn2009,Panasenco2011,Kliem2012}. 

The formation, eruption and rotation of the MFR are accompanied by the motion of 
footpoint brightenings, which has been comprehensively studied in a few decades. 
\citet{Su2006} shows that the footpoint brightenings observed by TRACE widely separate 
along the PIL in the initial stage, and then move away from the PIL gradually during the 
impulsive phase. A statistical study of footpoint motion of 50 X and M class two-ribbon 
flares indicates that both shear motion of conjugate footpoints and ribbon separation 
are common features in two-ribbon flares \citep{Su2007}. Following the eruption, 
post-flare loops and their footpoints propagate along the PIL are also 
observed, and the separation of the flare ribbons perpendicular to the PIL occurs at the 
same time or immediately after that \citep{Tripathi2006,Li2009}. In addition, 
hard X-ray (HXR) kernel motions parallel and perpendicular to the PIL have also been 
reported in previous studies \citep{Krucker2005,Liu2006,Yang2009}. \citet{Qiu2009} terms 
the two distinct stages of the flare ribbons evolution as stages of ``parallel 
elongation" and ``perpendicular expansion", which can be well explained by the two 
stages of three-dimensional (3D) reconnection of the erupted flux ropes (3D ``zipper 
reconnection" and quasi-2D ``main phase" reconnection) proposed by \citet{Priest2017}. 

The M6.5 class flare occurred on 2015 June 22 in active region NOAA 12371 is a 
complex flare which contains many physical processes including the tether-cutting 
reconnection, DAI and TI as suggested by \citet{Kang2019}. It 
has been the subject of numerous studies. Main progresses include the sudden 
flare-induced rotation of a sunspot and the association with the back reaction of the 
flare-related restructuring of coronal magnetic field \citep{Liu2016a}, the rotational 
motions of the photospheric magnetic flux and shear flows \citep{Bi2017,Wang2018a}, the 
evidence of a large-scale, long-duration, slipping-type reconnection \citep{Jing2017},
flare-ribbon-related photospheric magnetic field changes and the first evidence of the 
HXR coronal channel \citep{Liu2018a,Sahu2020}. \citet{Wang2017} studies the two 
precursors of this flare, and finds the low-atmospheric precursor emissions are 
closely related to the onset of the main flare. 

In our previous paper (\citealt{Liu2022}, hereafter Paper 1), we have studied 
the footpoint rotation and writhe of the two hot channels in the M6.5 class 
flare on 2015 June 22. However, the formation mechanism of the hot channels and their 
relationship with the two flare precursors are still unclear. The causes of the 
footpoint rotation and writhe of the hot channels also need further investigation. 
Therefore, these questions will be addressed in this study. Data set is introduced in 
Section 2. In Section 3, we present the observations results. We carry out magnetic 
topology analysis in Section 4. We summarize major findings and discuss the results in 
Section 5.

\section{Data Set}

The data used in this study are mainly from the Solar Dynamics
Observatory (SDO, \citealt{Pesnell2012}). The Atmospheric Imaging Assembly (AIA, 
\citealt{Lemen2012}) onboard SDO can simultaneously provide full-disk observations in 
EUV and UV passbands with temporal cadence of 12 and 24 seconds respectively, and the 
pixel size is 0.$^{\prime\prime}$6. AIA observations in 131~\AA, 304~\AA, 1600~\AA~are 
used to understand the hot channels, filaments, bright kernels and flare ribbons 
in this region \citep{O'Dwyer2010}. The magnetograms are provided by the Helioseismic 
Magnetic Imager (HMI, \citealt{Schou2012}) aboard SDO with a spatial resolution of 
0.$^{\prime\prime}$5/pixel, and a cadence of 45 seconds for line of sight (LOS) 
magnetograms (hmi.M$\_$45s series) and 720 seconds for vector magnetograms 
(hmi.sharp$\_$cea$\_$720s series). During ~16:25–22:50 UT on June 22, 2015, the 
1.6-meter Goode Solar Telescope (GST, \citealt{Cao2010}) at the Big Bear Solar 
Observatory (BBSO) took observations of the NOAA AR 12371 under excellent 
seeing conditions. The Visible Imaging Spectrometer (VIS) observations in H$\alpha$ 
(6563~\AA) line center with the time cadence of 28 seconds and pixel size of 
0.$^{\prime\prime}$03 are used to study the fine-scale structures at the chromosphere in 
unprecedented detail \citep{Jing2016}. The soft X-ray (SXR) emission of the flare has 
been recorded by GOES.

\section{Observations}

\subsection{Event Overview} 

\begin{figure}
	\begin{interactive}{animation}{Figure1movie.mp4}
		\plotone{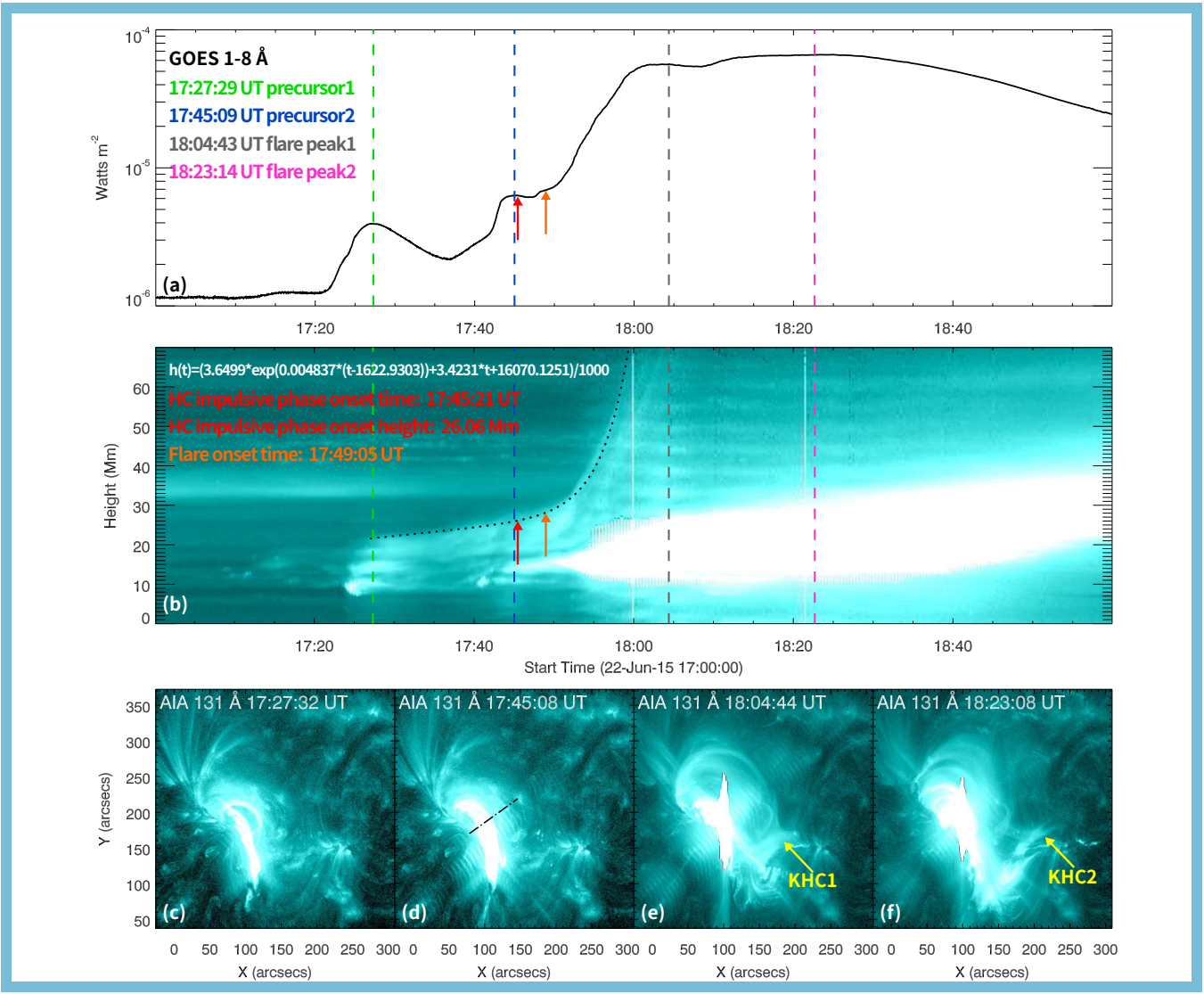}
	\end{interactive}
	\caption{Soft X-ray light curve and evolution of the hot channel. (a) 
		GOES soft X-ray light curve (1-8~\AA) between 17:00 and 19:00 UT on 2015 June 
		22. The middle image shows the time-distance diagram obtained from the stacking 
		images in 131~{\AA} by SDO/AIA at different times along the black dash dotted 
		line in panel (d). The green and blue vertical dashed lines in panels (a)-(b) 
		indicate the peak times of the first precursor and the second precursor.  The 
		gray and magenta vertical dashed lines mark the two peaks of the M6.5 class 
		flare. The red and orange arrows in panels (a)-(b) point out the onset times of 
		the seed hot channels' impulsive phase and the flare. The fit result for the 
	    time evolution of the height of hot channels is shown by the black dotted line 
	    in panel (b), and the fit function is on the upper left corner of the image. 
	    (c)-(f) AIA 131~{\AA} images at the peak times of the first flare precursor, the 
	    second flare precursor and the flare, respectively. The two yellow 
	    arrows in panels (e) and (f) mark the two kinking hot channels. An animation of 
	    this figure is available. It covers 2 hours of observation beginning at 17:00:08 
	    UT on 2015 Jun 22. The video duration is 24 seconds. 
		\label{fig:lightcurve}}
\end{figure}

A C1.1 class flare occurs in AR 12371, which is a confined flare without CME 
\citep{Awasthi2018}, and the peak time is 16:45 UT. Then two flare precursors are 
observed prior to the main phase of the M6.5 class eruptive flare 
\citep{Wang2017}. The peak times of the two flare precursors  
(17:27 UT and 17:45 UT) are marked by the green and blue vertical dashed lines in the 
GOES light curve shown in Figure \ref{fig:lightcurve}(a). During the first precursor, 
seed hot channels build up and rise slowly, being accelerated at the peak of the second 
precursor, as shown in Figures \ref{fig:lightcurve}(b)-(d).

The observed height-time plot of the hot channel along the erupting direction 
(marked as black dash-dotted line in Figure \ref{fig:lightcurve}(d)) is fitted by an 
analytic approximation with the combination of linear (slow rise) and exponential (fast 
rise) functions developed by \citet{Cheng2020}, which is shown as the black dotted line 
in Figure \ref{fig:lightcurve}(b). Because this active region is close to the solar 
disk, the estimation of the heights, velocities, and accelerations of the hot channels 
may be significantly influenced by the projection effect, but the characteristics of the 
temporal profile of the hot channels are not affected \citep{Cheng2020}. The fitting 
results show that the onset time of impulsive acceleration phase is 17:45:21 UT ($\pm 
1$ m, marked by the red arrows in Figures \ref{fig:lightcurve}(a)-(b)). A comparison 
of Figures \ref{fig:lightcurve}(a) and \ref{fig:lightcurve}(b) shows that the onset of 
the impulsive acceleration of the seed hot channels is earlier than the flare onset 
(marked by the orange arrows in Figures \ref{fig:lightcurve}(a)-(b)). 
After that, the seed hot channels rise rapidly while the filaments in AIA 304~\AA~ and 
VIS H$\alpha$ images stay behind. After the flare onset, two hot channels form 
one after another (marked by the yellow arrows in Figures 
\ref{fig:lightcurve}(e)-(f)), and they both exhibit a kinking structure with negative 
crossing. The eruption of these two hot channels produce two peaks (marked by the gray 
and magenta vertical dashed lines in Figure \ref{fig:lightcurve}(a)) on the flare's GOES 
light curve. The morphological evolution and footpoint motion of these two hot channels 
are studied in Paper 1. 

\subsection{Formation and Buildup of the Seed Hot Channels}

\begin{figure}
	\plotone{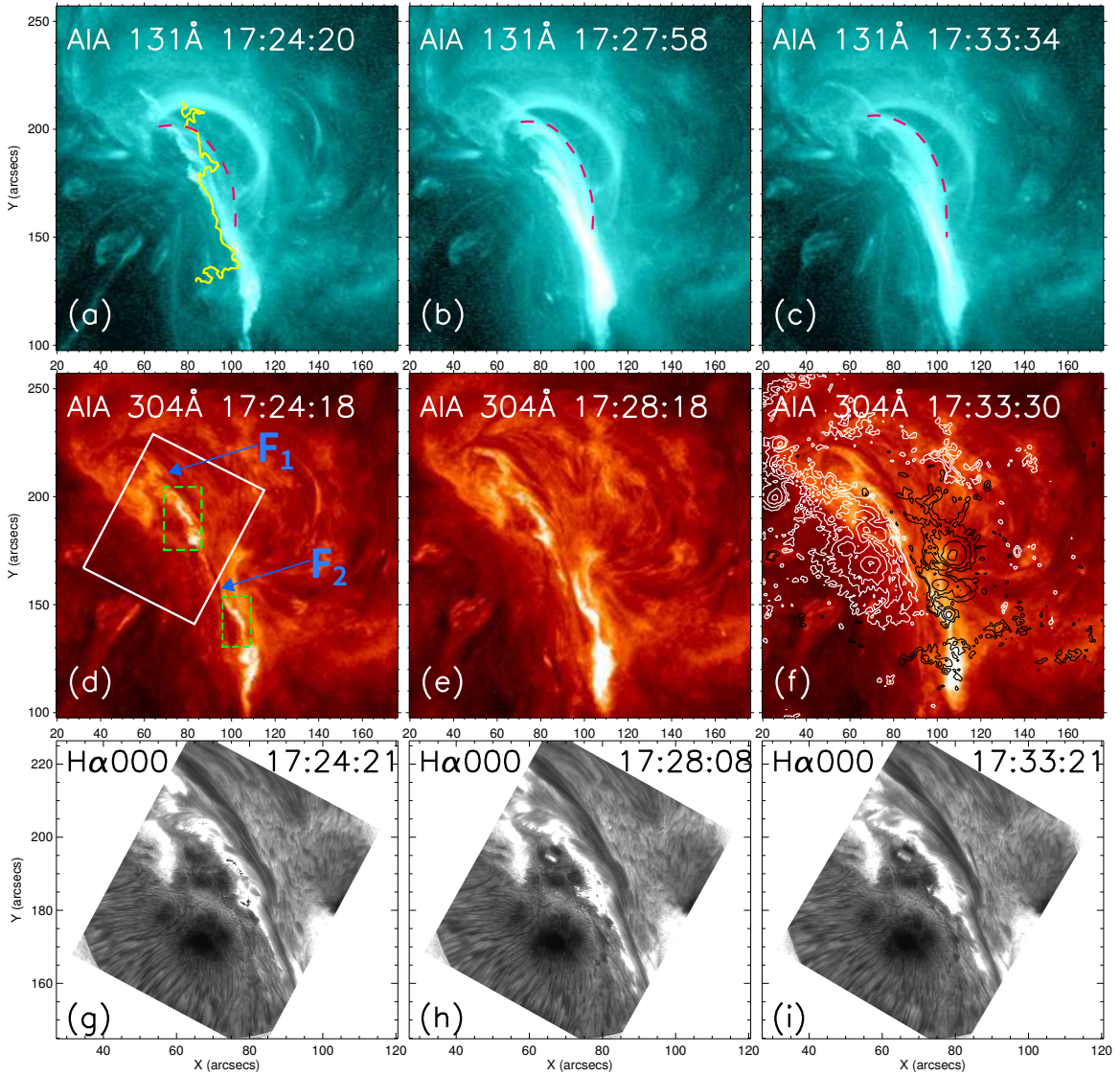}
	\caption{Morphological evolution at the first flare precursor. (a)-(i) 
		Multi-wavelength images acquired by SDO/AIA 131~\AA~and 304~\AA, BBSO/GST 
		H$\alpha$ center line. The magenta dashed lines in panels (a)-(c) draw the outer 
		edge of the seed hot channels during the first precursor. The yellow curve in 
		panel (a) marks the main PIL of the active region. The white box in 
		panel (d) represents the FOV of panels (g)-(i), the blue arrows mark the two 
		filaments observed in AIA 304~\AA, and the green dashed boxes frame the 
		brightenings that appear on both sides of the PIL. The white and black contours 
		in panel (f) show the photospheric positive and negative magnetic fields taken 
		by SDO/HMI at 17:33:30 UT. 
		\label{fig:precursor1}}
\end{figure}

\begin{figure}
	\plotone{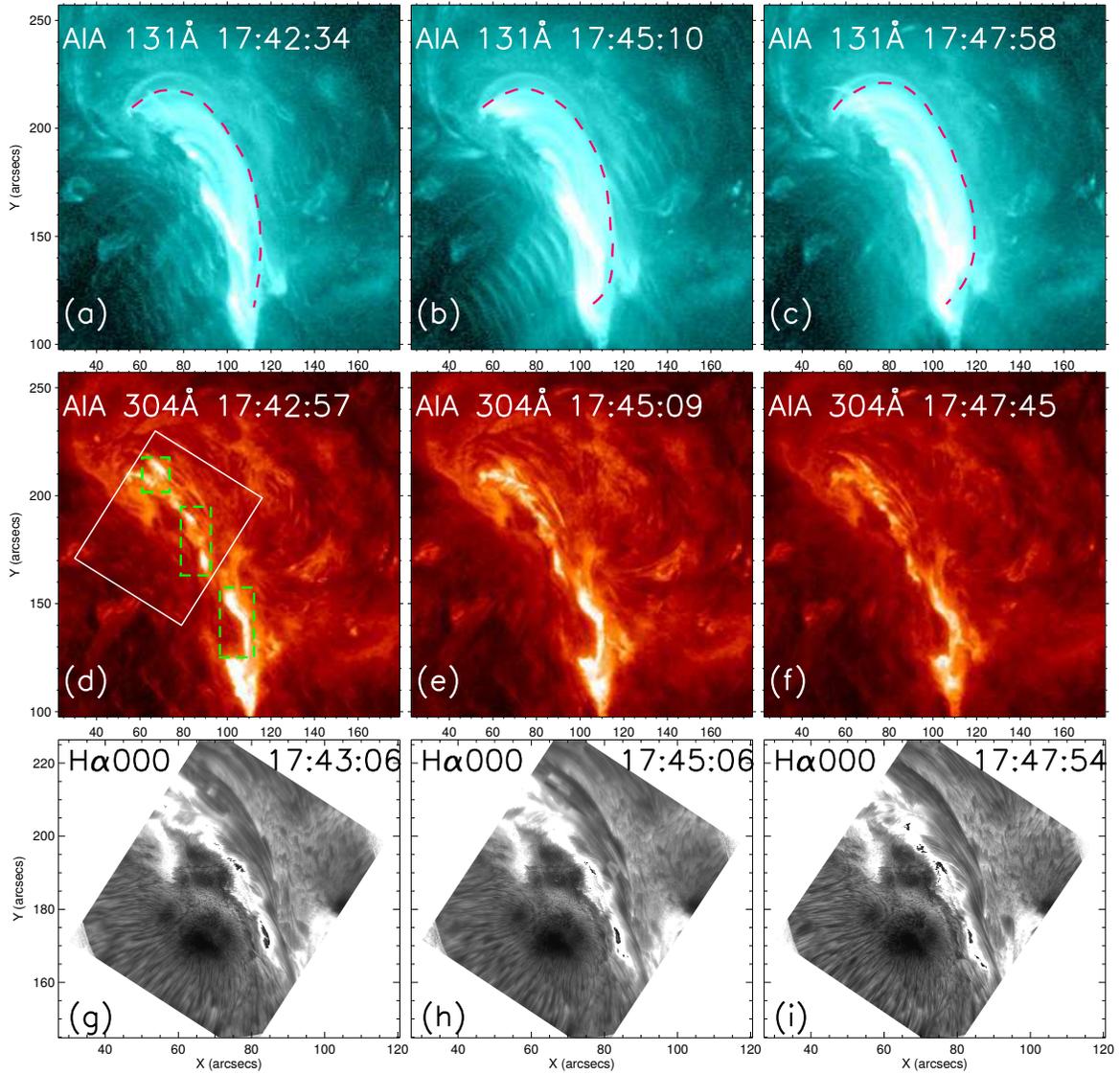}
	\caption{Morphological evolution at the second flare precursor. (a)-(i) 
		Multi-wavelength images acquired by SDO/AIA 131~\AA~and 304~\AA, BBSO/GST 
		H$\alpha$ center line. The magenta dashed lines in panels (a)-(c) draw the outer 
		edge of the seed hot channels during the second precursor. The white box 
		in panel (d) represents the FOV of panels (g)-(i). The green dashed boxes frame 
		the brightenings that appear on both sides of the PIL.
		\label{fig:precursor2}}
\end{figure}

After a high-resolution investigation of the two flare precursors before the M6.5 
class flare, \citet{Wang2017} has concluded that the eruption of the main flare is 
resulted from the successive reconnection between the sheared loops. In this study we 
focus on the relationship between the flare precursors and the hot channels. The 
morphological evolution of the two successive episodes of precursors observed by AIA/SDO 
and VIS/GST is presented in Figures \ref{fig:precursor1}-\ref{fig:precursor2}. At the 
beginning of these two precursors, brightenings appear on the two sides of the PIL at 
around 17:24 UT (marked by the green boxes in Figure \ref{fig:precursor1}(d)) 
and 17:42 UT (marked by the green boxes in Figure \ref{fig:precursor2}(d)) 
respectively. Starting from 17:24 UT, a few hot loops are observed in 131~\AA~ 
by AIA (Figures \ref{fig:precursor1}(a)-(c)), which are the early morphology of the hot 
channel, thus termed as seed hot channels. These seed hot channels are manifested as a 
group of brightened branches in high-temperature passbands, and the magenta dashed lines 
in Figures \ref{fig:precursor1}(a)-(c) mark the outer edge of them. As more and more hot 
branches brighten, the shape of the seed hot channels become clear. Two filaments can be 
identified in the corresponding AIA 304~\AA~images (pointed by the blue arrows in Figure 
\ref{fig:precursor1}(d)) and the VIS/GST H$\alpha$ line center images (Figures 
\ref{fig:precursor1}(g)-(i)), the field of view (FOV) of which is shown in the white 
box in Figure \ref{fig:precursor1}(d). The shape and location of these two filaments 
remain unchanged during the first flare precursor.

A comparison of Figure \ref{fig:precursor1} and Figure \ref{fig:precursor2} shows that, 
the brightenings in AIA 131~\AA, 304~\AA, and VIS H$\alpha$ images (Figure 
\ref{fig:precursor2}) during the second flare precursor are brighter than those in the 
first precursor. At the second precursor, the outer edge of the seed hot channels 
(marked by the magenta dashed lines in Figures \ref{fig:precursor2}(a)-(c)) 
becomes higher and longer, and the two end points almost connect the two distant ends of 
the two filaments. As the surrounding brightenings increase, the filaments cannot be 
clearly recognized in 304~\AA~ by AIA. As shown in the corresponding H$\alpha$ 
images in Figures \ref{fig:precursor2}(g)-(i), brightenings and motions of 
filament materials are identified and the filament becomes wider, which is similar to 
the seed hot channels. 

The seed hot channels appear during the first flare precursor, and becomes 
significantly larger after the second flare precursor. Their footpoint 
brightenings are observed at both sides of the PIL during the precursors which is 
different from the flare ribbons observed during the flare main phase. As we mentioned 
before, the outer edge of the seed hot channels doesn't change obviously during 
the first precursor, and the two footpoints only extend for about 2$^{\prime\prime}$ 
(Figures \ref{fig:precursor1}(a)-(c)). Before the onset of the second precursor, the 
northern footpoints of the seed hot channels continue to move northward for 
2$^{\prime\prime}$, and the southern footpoints expand southward by about 
38$^{\prime\prime}$, then the seed hot channels are clearly larger. The 
locations of the two footpoints hardly change during the second precursor, but the width 
of the seed hot channels increases significantly. In other words, the 
observations indicate a buildup process of the seed hot channels from the 
first precursor to the second precursor. 

\subsection{Propagation of the Footpoint Brightenings }

\begin{figure}
\begin{interactive}{animation}{movie1.mp4}
	\plotone{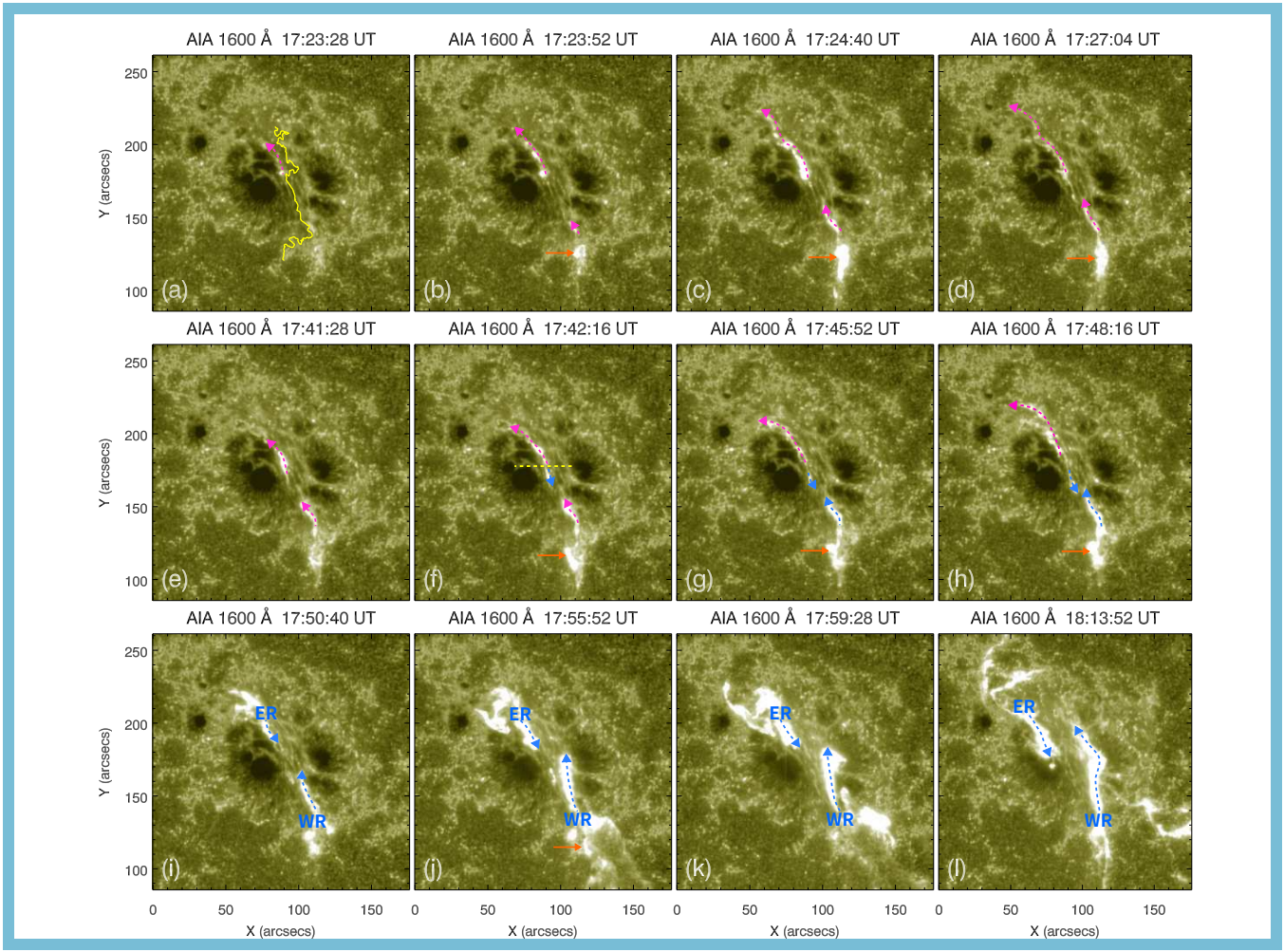}
\end{interactive}
	\caption{The propagation of bright kernels during the flare precursors and 
	flare main phase. (a)-(d) Propagation of bright kernels during the first precursor. 
    The magenta dotted arrows indicate the position of the brightening distribution at 
    the corresponding time. The yellow curve in panel (a) represents the PIL. The orange 
    arrows in panels (b)-(d) represent the brightening at the southern end of the west 
    bright ribbon and the first jet evolved from it. (e)-(h) Propagation of bright 
    kernels during the second precursor. The magenta and blue dotted arrows indicate the 
    position of the brightening distribution at the corresponding time, where the 
    magenta dotted arrows represent the brightening with northward motion, and the blue 
    dotted arrows represent the brightening with converging motion. The yellow dashed 
    line in panel (f) represents the dividing line of the northern and southern parts of 
    the east bright ribbon, and the orange arrows in panels (f)-(h) represent the 
    brightening at the southern end of the west bright ribbon and the second jet evolved 
    from it. (i)-(l) Propagation of bright kernels during the flare main phase. The blue 
    dotted arrows ER and WR in panels (i)-(l) represent the east and west flare ribbons, 
    respectively. The orange arrow in panel (j) indicates the third jet. An animation of 
    this figure is available. It covers 2 hours of observing beginning at 16:59:52 UT on 
    2015 Jun 22. The video duration is 24 seconds.
	\label{fig:bright}}
\end{figure}

In this subsection, we focus on the propagation direction of the brightenings during the 
two precursors and the flare main phase. The bright kernels in AIA 1600~\AA~are shown 
in Figure \ref{fig:bright}. At the onset of the first precursor, the bright kernels 
first appear at the east side of the PIL, and they are tracked by the magenta 
dashed arrow in Figure \ref{fig:bright}(a), and the bright kernels gradually move 
northward (see Figures \ref{fig:bright}(a)-(b)). Subsequently, a bright kernel 
appears on the west side of the PIL, which also moves northward parallel to the PIL (see 
Figures \ref{fig:bright}(b)-(d)). At the west side of the PIL another bright kernel 
appears at the south end and gradually extends southward, then evolves into a southward 
jet (marked by orange arrows in Figures \ref{fig:bright}(b)-(d)), while the northward 
propagating bright kernels gradually disappear at the end of the first flare precursor.

At the beginning of the second flare precursor, bright kernels reappear on both sides of 
the PIL and are more widely distributed than those at the beginning of 
the first flare precursor (see Figure \ref{fig:bright}(e)). Immediately, these bright 
kernels expand and spread toward the north (see Figure \ref{fig:bright}(f)). 
Next, the west bright ribbon almost stops moving while the east bright ribbon splits 
into two parts (northern and southern parts, separated by the yellow dashed line 
in Figure \ref{fig:bright}(g)) that move northward and southward, respectively 
(see Figures \ref{fig:bright}(f)-(g)). And the southern part of the east bright ribbon 
almost stops moving after two minutes. After the peak of the second flare precursor, the 
west bright ribbon and the southern part of the east bright ribbon start converging 
along the PIL. The northern part of the east bright ribbon keeps moving northward 
during this process (see Figures \ref{fig:bright}(f)-(h)). During the second flare 
precursor, there is also a bright kernel at the southern end of the west bright ribbon, 
which gradually expands and then evolves into a southward jet (marked by orange arrows 
in Figures \ref{fig:bright}(f)-(h)).

After the beginning of the flare main phase, the southern part of the east bright 
ribbon disappears, and the northern part begins to strengthen significantly, forming a 
flare ribbon (labeled `ER' in Figures \ref{fig:bright}(i)-(l)) and gradually moving 
southward. Another flare ribbon (labeled `WR' in Figures \ref{fig:bright}(i)-(l)) 
evolved from the west bright ribbon moves northward obviously. The subsequent southward 
jet from the western flare ribbon is more intense than those during the two flare 
precursors (marked by orange arrow in Figure \ref{fig:bright}(j)). From about 
17:56:40 UT, the anti-parallel motion of the two flare ribbons begins to accompany the 
separation motion, which lasts about ten minutes. Then, with the appearance of flare 
loops, the two flare ribbons only move away from each other in the direction 
perpendicular to the PIL (see Figures \ref{fig:bright}(k)-(l)).   

In general, the bright kernels on both sides of the PIL mainly show northward 
parallel motion during the first precursor. The brightenings show prominent parallel 
motion toward the north and converging motion along the PIL during the second 
precursor. During the flare main phase, the brightenings first display converging motion 
along the PIL, and then expand with converging along the PIL, finally move perpendicular 
to the PIL.

\subsection{Formation and Evolution of the Two Hot Channels}

\begin{figure}
	\plotone{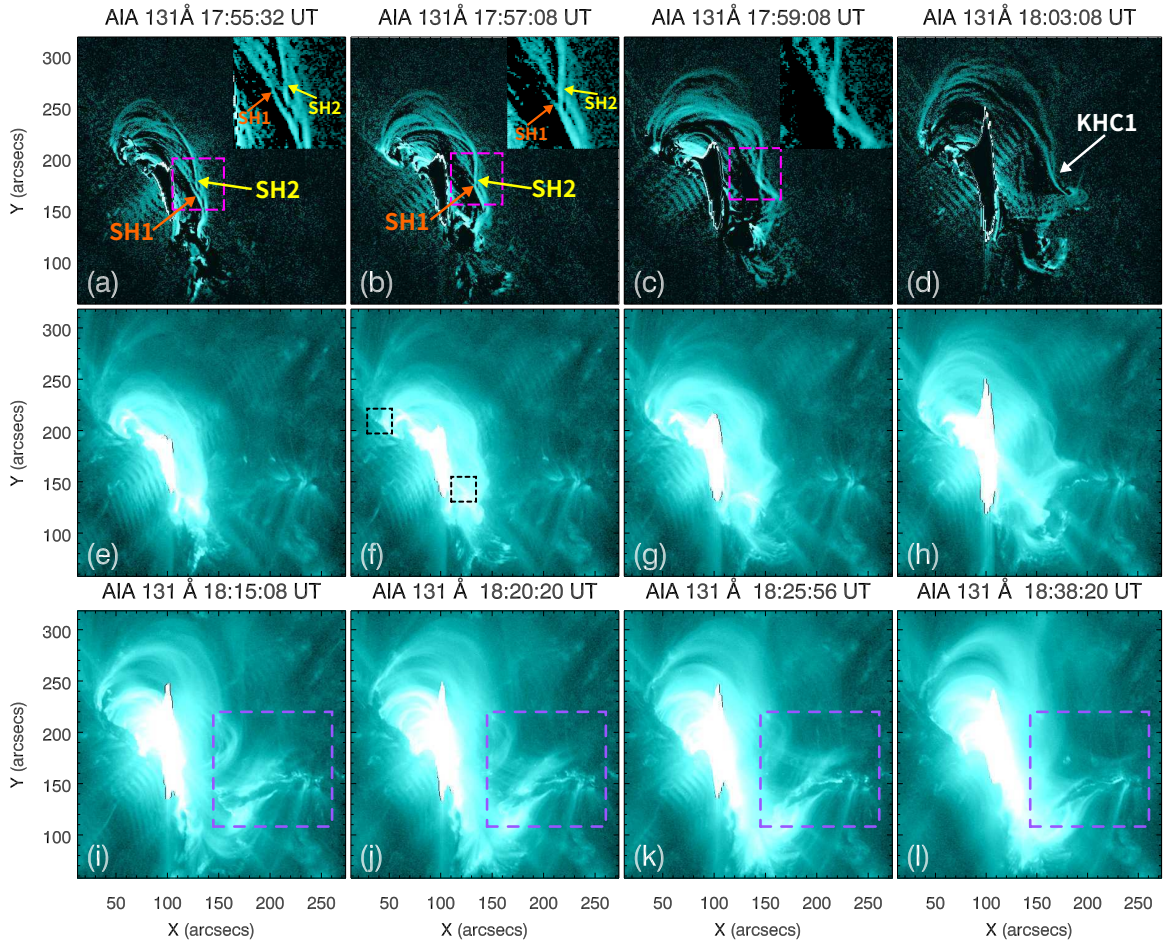}
	\caption{Kinking evolution of hot channels. The first row shows the running 
	difference images in the AIA 131~\AA~during the evolution of hot channels. Arrows 
	SH1 (orange) and SH2 (yellow) represent two seed hot channels that merging. 
	The insets at the upper right corner of panels (a)-(c) are enlarged portion 
	of the region framed by the magenta boxes in corresponding panels, to better show 
	the merge of the two seed hot channels. The white arrow KHC1 in panel (d) represents 
	the first kinking hot channel formed by merging. The second row presents the 
    original images observed in AIA 131~\AA~at the same time as those at the first row. 
    The black boxes in panel (f) mark footpoints brightening of the first kinking hot 
    channel. The third row shows the original images observed in AIA 131~\AA~during the 
    evolution of second kinking hot channel, the purple boxes in panels (i)-(l) mark the 
    kinking structure of this hot channel.  
	\label{fig:reconnection}}
\end{figure}

After the onset of the impulsive phase, the seed hot channels expand 
rapidly. The AIA images in 131~\AA~show that the seed hot channels consist of multiple 
branches, which can be seen more clearly from the running-difference images 
(see Figure \ref{fig:reconnection}). Parallel flux tubes with the same twist can merge 
into a single flux tube near the point of contact \citep{Linton2001}. 
At 17:55:32 UT, two seed hot channels marked by the orange and yellow arrows (labeled 
`SH1', `SH2') in Figure \ref{fig:reconnection}(a) are identified. These two seed hot 
channels are close together, but there is a clear gap between them (see Figure 
\ref{fig:reconnection}(a) and its inset). With the expansion of the hot channel, SH1 
and SH2 gradually intersect (see Figures \ref{fig:reconnection}(b)-(c)), 
accompanied by the appearance of footpoint brightenings of the seed hot channels (marked 
by the black boxes in Figure \ref{fig:reconnection}(f)), which suggest the occurrence of 
merge reconnection. In the process of impulsive acceleration, the different branches of 
the seed hot channels continuously merge, and at 17:59:08 UT, a longer and more twisted 
hot channel (i.e., the first kinking hot channel, labeled `KHC1' in Figure 
\ref{fig:reconnection}(d)) forms, which subsequently appears as a kinking structure (see 
Figure \ref{fig:reconnection}(d)). About three minutes later, the kinking structure 
disappears and both footpoints of the first kinking hot channel display an apparent 
clockwise rotation during the unwrithing of this hot channel, which has been studied in 
detail in Paper 1.

During the unwrithing of the first kinking hot channel (KHC1), another hot channel 
(KHC2) appears near the right leg of KHC1 with a kinking structure (marked by the purple 
box in Figure \ref{fig:reconnection}(i)), which can only be barely  
distinguished at this time due to the envelope of KHC1's leg. By 18:20:20 UT, the 
footpoint rotation of KHC1 ends, the kinking structure of KHC2 can be clearly observed, 
and obvious brightening can be seen at its right footpoint 
(see Figure \ref{fig:reconnection}(j)). Immediately, the left and right footpoints of 
KHC2 begin to move northward and westward, respectively. The movement of the right 
footpoint is more obviously, and the KHC2 gradually unwrithes as the right footpoint 
slides to the west (see Figures \ref{fig:reconnection}(j)-(k)). By 18:38:20 UT, the 
kinking structure almost disappears, and only the brightening at the right footpoint can 
be seen (see Figure \ref{fig:reconnection}(l)).

\section{Magnetic Field Modeling}

To understand the 3D topology of the source regions of this events, we analyze the 
magnetic field characteristics of this active region based on the nonlinear  
force-free field (NLFFF) extrapolations by \citet{Awasthi2018}. The vector 
magnetograms at different times are remapped at the original resolution by the Lambert 
(cylindrical equal area; CEA) projection method. Then a “pre-processing" procedure is 
used to remove the net force and torque of the photospheric field to best fit the 
force-free condition \citep{Wiegelmann2006}. Finally, these “preprocessed” magnetograms 
are input into the NLFFF code proposed by \citet{Wiegelmann2004} as boundary conditions, 
and the “weighted optimization" method is applied to obtain the time series of the NLFFF 
models. The MFR can be identified through mapping magnetic connectivities and computing 
the twist number ($T_{w}$) for each individual field line \citep{Liu2016b}. $T_{w}$ 
measures the number of turns of a field line winding, and it is calculated by 
integrating the local density of 
$T_{w}$, $\nabla\times\boldsymbol{B}\cdot\boldsymbol{B}/4\pi B^{2}$, along each field 
line \citep{Awasthi2018, Liu2018a}. 

\subsection{Twist Evolution of the Flux Ropes in the NLFFF Extrapolations}

\begin{figure}
	\plotone{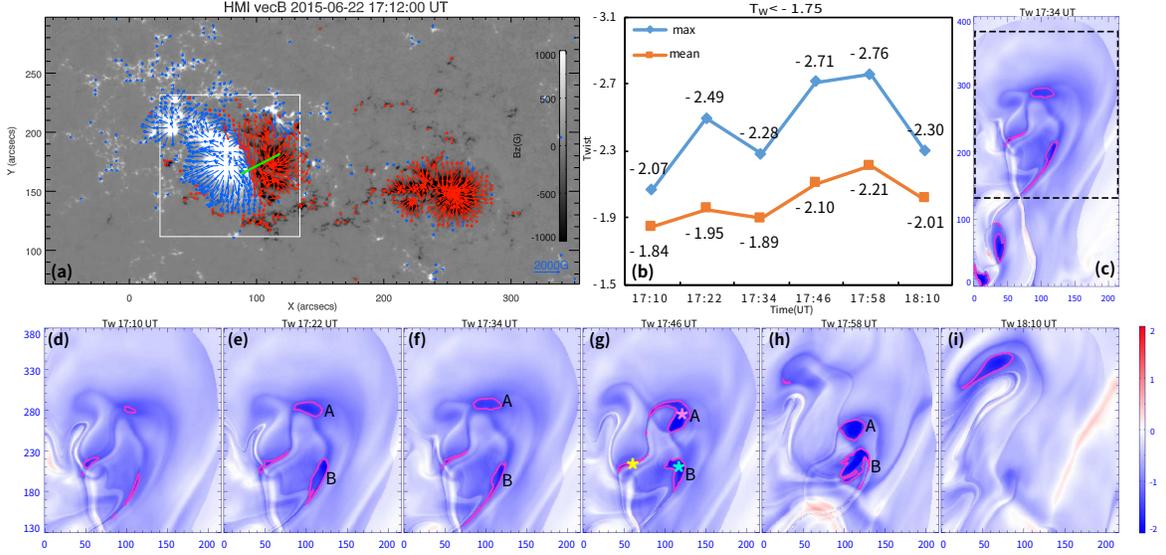}
	\caption{(a) HMI photospheric vector magnetogram at 17:12 UT. The background map is 
	the vertical component of the magnetic field, the intensity of which is represented 
	by the color bar on the right. The blue (positive) and red (negative) arrows 
	represent the horizontal magnetic fields, and the strength of which is 
	represented by the length of the arrows. The white rectangle represents the 
	FOV of images at the second row of Figure \ref{fig:flux rope}. (b) Temporal 
	evolution of the max (blue) and mean (orange) $T_{w}$ in the regions with $T_{w} \le 
	-1.75$. (c)-(i) $T_{w}$ distributions in the X-Z cross sections along the green line 
	shown in panel (a). The black dotted box in panel (c) frames the FOV of 
	panels (d)-(i). The ranges of Z are 0$^{\prime\prime}-50^{\prime\prime}$ for 
	panel (c) and 16$^{\prime\prime}$.25$-$47$^{\prime\prime}$.5 for panels (d)-(i). The 
	labels `A' and `B' mark the upper and right regions with $T_{w} \le -1.75$. 
	The cyan, pink and yellow `*' symbols in panel (g) indicate the flux ropes of the 
	same color code in Figure \ref{fig:flux rope}. The magenta contours in 
    panels (c)-(i) represent the $T_{w}$ of $-1.75$.  
	\label{fig:twist}}
\end{figure}

Using the method proposed by \citet{Liu2016b}, we calculate the distribution of twist 
number in this active region. The photospheric vector magnetogram at 17:12 UT is 
presented in Figure \ref{fig:twist}(a). The cross sections of the $T_{w}$ maps in the 
X-Z plane along the green line marked in Figure \ref{fig:twist}(a) are presented in 
Figures \ref{fig:twist}(c)-(i), in which the contour with $T_{w} = -1.75$ is outlined in 
magenta. At 17:10 UT, the  absolute $T_{w}$ values in three regions exceed 1.75 (see 
Figure \ref{fig:twist}(d)). With the beginning of flare precursors, the area of these 
three regions increases, and the increase of the upper (marked as `A') and right (marked 
as `B') regions are more obvious (see Figures \ref{fig:twist}(e)-(f)). At the later 
stage of the second flare precursor (Figure \ref{fig:twist}(g)), the left region becomes 
slender, the area of the upper and right regions increases. Next, the upper and right 
regions gradually approach. At 18:10 UT, there is only one larger area with 
$T_{w} \le -1.75$ on the left (see Figures \ref{fig:twist}(h)-(i)). 

Based on the coronal magnetic field obtained by NLFFF extrapolation, we calculate the 
variation of mean and maximum values of $T_{w}$ over time in the area 
where $T_{w}$ is lower than $-1.75$ in Figures \ref{fig:twist}(d)-(i), and the 
calculation results are shown in Figure \ref{fig:twist}(b). The calculated maximum and 
mean $T_{w}$ increase during the flare precursors (except during 17:22 UT-17:34 UT), 
reach the maximum at 17:58 UT, and then begin to decline. 

\begin{figure}
	\plotone{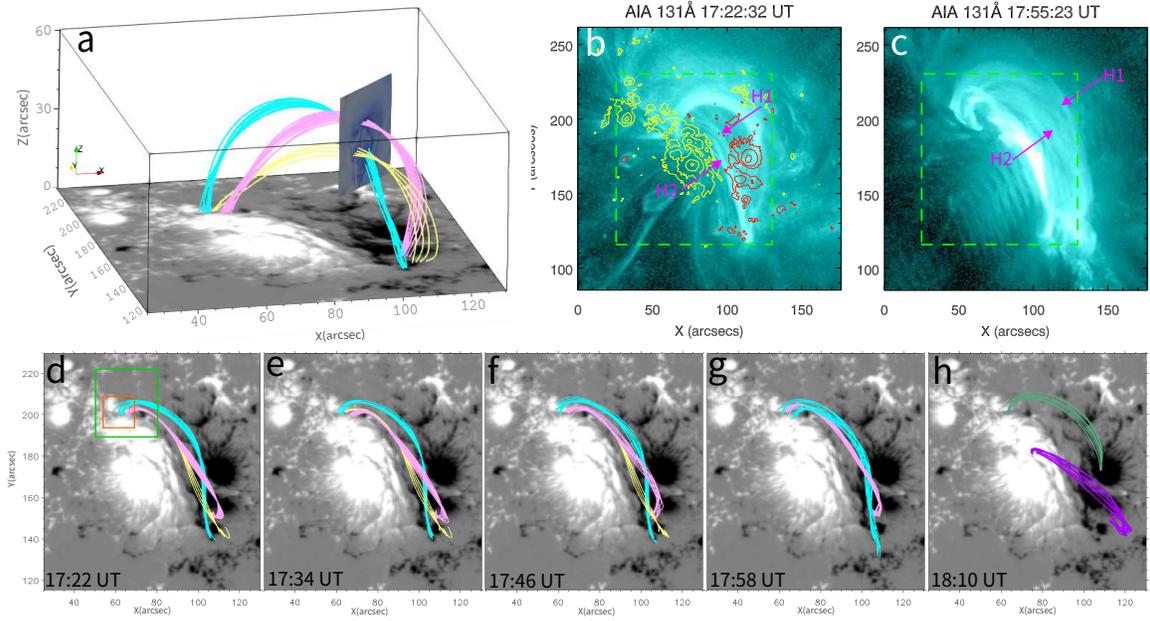}
	\caption{(a) Magnetic field lines derived from the NLFFF at 17:34:25 UT in 3D 
		perspectives, the cross section of X-Z plane is the same as those in Figure 
		\ref{fig:twist}(f). The cyan, pink and yellow field lines represent the flux 
		ropes passing through the three regions with $T_{w} \le -1.75$. The 
		FOV of images in panels (d)-(h) is marked by the white box in Figure 
		\ref{fig:twist}(a). (b)-(c) Images in AIA 131~\AA~at 17:22:32 UT and 
		17:55:23 UT, and the green boxes represent the FOV of images in panels (d)-(h), 
		the magenta arrows H1 and H2 represent the two observed hot channels. The 
		yellow and red contours in panel (b) show the photospheric positive and negative 
		magnetic fields taken by SDO/HMI at 17:22:30 UT. (d)-(h) Magnetic field lines 
		superimposed on an HMI LOS images at 17:22:25 UT, 17:34:25 UT, 17:46:25 UT, 
		17:58:25 UT and 18:10:25 UT, respectively. The orange box in panel (d) 
		represents the FOV of images in Figure \ref{fig:left footpoint}. The 
		green and purple lines in panel (h) represent the newly formed flux bundles and 
		a new flux rope respectively.
		\label{fig:flux rope}}
\end{figure}

We trace the magnetic field lines passing through these regions with high $T_{w}$ at 
different times, and the results are presented in Figure \ref{fig:flux rope}. A 3D view 
of selected magnetic field lines of the NLFFF at 17:34 UT is shown in Figure 
\ref{fig:flux rope}(a), the $T_{w}$ map in the X-Z plane is also shown in this 
panel. Yellow, pink and cyan lines represent the flux ropes passing through the strong 
$T_{w}$ regions on the left, upper and right sides. We select the central area of the 
active region (marked by the white box in Figure \ref{fig:twist}(a)) to study the 
evolution of these twisted flux ropes. The flux ropes represented by the same color are 
displayed with the same resolution at different times. From 17:22 UT to 17:58 UT, the 
projections of pink and cyan flux ropes on the X-Y plane are getting closer and 
closer, while the yellow flux rope disappears at 17:58 UT. The shape and relative 
positions of the cyan and pink flux ropes at 17:22 UT and 17:58 UT are similar to the 
seed hot channels H1 and H2 (marked by magenta arrows in 
Figures \ref{fig:flux rope}(b)-(c)) observed in AIA 131~\AA~ at the 
corresponding time. The FOV of images in panels (d)-(h) is 
marked by the green boxes in panels (b) and (c). It can be seen that the height and 
length of the extrapolated magnetic field lines are smaller than those of the observed 
hot channels. This may be caused by uncertainties of the magnetic field measurements or 
over-smoothing of the vector field by the extrapolation preprocessing. Because the 
flux ropes during the eruption process is under Lorentz force, it is difficult to be 
well reproduced by NLFFF \citep{Cheng2016}. At 18:10 UT, only one new twisted flux rope 
can be identified (purple line in Figure \ref{fig:flux rope}(h)), and its X-Z cross 
section corresponds to the area surrounded by the magenta contour in 
Figure \ref{fig:twist}(i). The right footpoint of this new flux rope is near the right 
footpoint of the disappeared pink flux rope, but the left footpoint is far away from the 
left footpoint of the original flux rope.

\begin{figure}
	\plotone{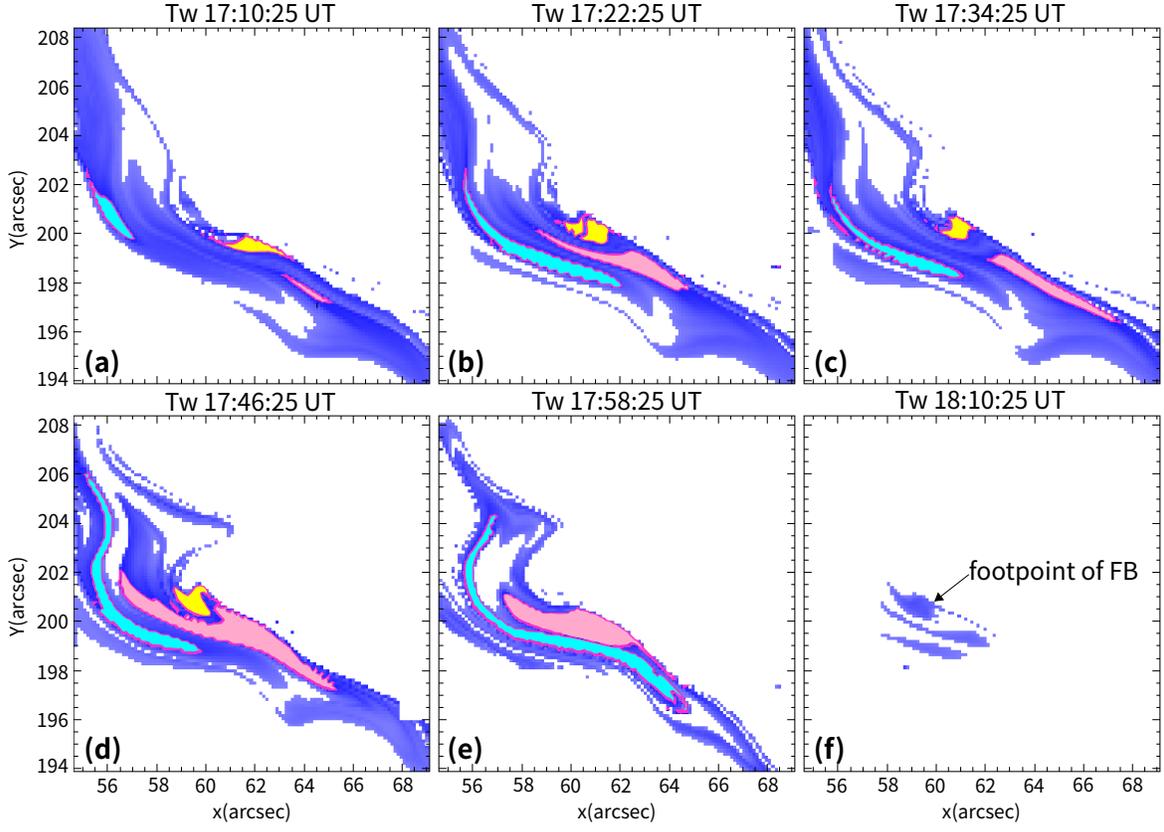}
	\caption{Evolution of $T_{w}$ distribution at the left footpoint of flux ropes. 
	To better show the evolution, only the twist distribution at the location of 
	$T_{w} \le -1$ are shown, the twist values at other locations are set to zero. The 
	magenta contours represent the $T_{w}$ at -1.75. The cyan, pink and yellow fills 
	correspond to the footpoints of the original three flux ropes, respectively. The 
    arrow in panel (f) marks the footpoint of the newly appeared flux bundles.  
	\label{fig:left footpoint}}
\end{figure}

In order to study the footpoint motion of these flux ropes during the flare, we 
select the area surrounded by the orange box in Figure \ref{fig:flux rope}(d) to show the
temporal evolution of the left footpoint of these flux ropes, and the results are 
presented in Figure \ref{fig:left footpoint}. To better show the evolution, only the 
twist distribution at the locations with $T_{w} \le -1$ is shown, and the twist values 
at other locations are set to zero. The magenta contours represent the $T_{w}$ at 
-1.75. The footpoints of the cyan, yellow and pink flux ropes are filled with 
corresponding colors respectively. At 17:10:25 UT, these three flux ropes are 
separated from each other (see Figure \ref{fig:left footpoint}(a)). The footpoint areas 
(especially the cyan and pink areas) increase significantly with the beginning of the 
first precursor (see Figure \ref{fig:left footpoint}(b)), so they appear to close to 
each other. At the end of the first precursor (see Figure \ref{fig:left footpoint}(c)), 
the footpoints seem to separate from each other again, which may be caused by the 
decrease of the yellow areas and the deformation of the pink areas. During the second 
precursor, with the significantly increase of the cyan and pink areas (see Figure 
\ref{fig:left footpoint}(d)), the three footpoints gather together again. The yellow 
footpoint is surrounded by the pink footpoint at 17:46:25 UT. By 17:58:25 UT, the yellow 
footpoint area disappears, the cyan and pink footpoint areas connect together. At 
18:10:25 UT, the footpoints of the original flux ropes disappear (see Figure 
\ref{fig:left footpoint}(f)), and a flux bundle represented by the green lines (see 
Figure \ref{fig:flux rope}(f)) appears with the same left footpoint position as the 
original flux rope and smaller twist. 

\subsection{Torus Instability of the Extrapolated Flux ropes}

As we mentioned before, the onset of the fast rise of the seed hot channels 
is earlier than the flare onset, which suggests that the initiation of 
the impulsive acceleration of the seed hot channels is unlikely caused by 
flare reconnection. The torus instability occurs when the inward 
tension force generated by the background magnetic field decreases faster than 
the outward hoop force \citep{Kliem&Torok2006,Fan2007}. TI is quantified by the decay 
index n, which is defined by $n = -\frac{\mathrm{d}\ln{B}}{\mathrm{d}\ln{z}}$. Here, B 
denotes the background magnetic field strength and z denotes the height above 
the solar surface \citep{Cheng2013b}. Previous studies suggest that the threshold value 
of decal index is lie in a range of 1.1-2, and normally 1.5 for a toroidal flux rope
\citep{Torok&Kliem2005, Kliem&Torok2006, Aulanier2010}. To investigate the initiation 
mechanism of the impulsive acceleration of the seed hot channels, we calculate the decay 
index of the background magnetic field, the background field is obtained from the 
potential field model. Since the vertical component does not contribute to the downward 
constraining force applied to the flux rope, only the horizontal component of the 
coronal magnetic field is considered in the calculation of the decay index 
\citep{Cheng2013a}.

\begin{figure}
	\plotone{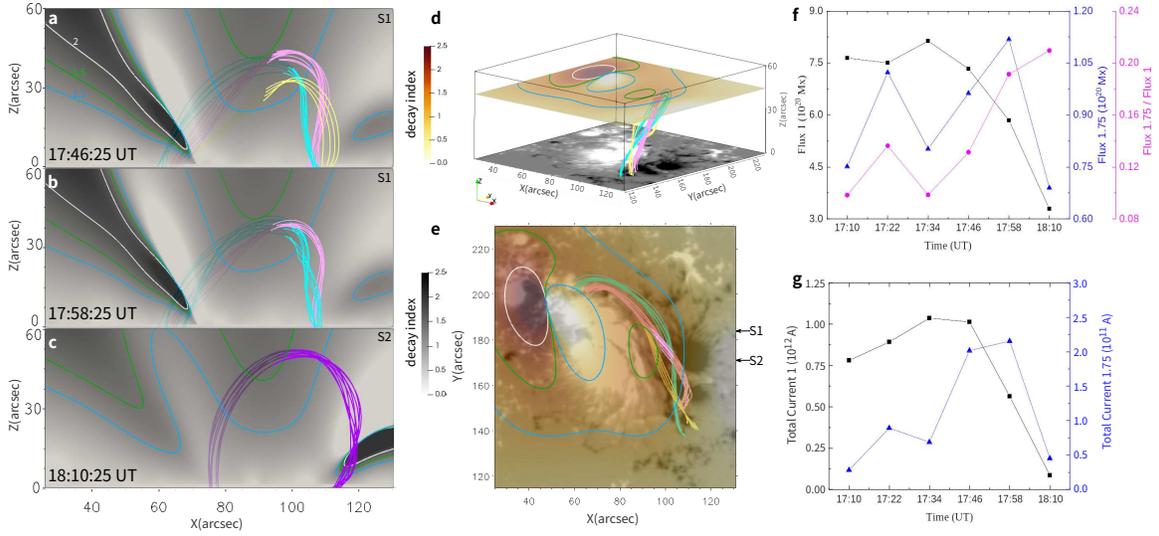}
	\caption{(a)-(c) Decay index distribution in X-Z planes at three different moments. 
	The cyan, pink, yellow and purple lines refer to the same field lines as those in 
	Figure \ref{fig:flux rope}. The blue, green, white contours in panels 
	(a)-(e) refer to the decay index at 1.1, 1.5 and 2, respectively. (d) NLFFF lines 
    superimposed on an HMI LOS images at 17:46:25 UT in 3D perspectives, the orange 
    cross section in X-Y plane represents the distribution of decay index above the top 
    of the flux ropes. (e) Top view of image in panel (d), the arrow S1 marks the 
    position of images in panels (a)-(b) and the arrow S2 marks the position of image in 
    panel (c). (f) The evolution of magnetic flux in the areas where $T_{w} \le -1$ 
    (black) and $T_{w} \le -1.75$ (blue) in the region marked by the green box in 
    Figure \ref{fig:flux rope}(d), and the evolution of the magnetic flux ratio 
    (magenta) in the areas with $T_{w} \le -1.75$ and $T_{w} \le -1$. (g) The evolution 
    of total current of the flux ropes satisfying $T_{w} \le -1.75$ (blue) and 
    $T_{w} \le -1$ (black) in the region marked by the green box in 
    Figure \ref{fig:flux rope}(d). 
	\label{fig:decay index}}
\end{figure}

The relationship between the position of the magnetic flux ropes and the distribution of 
decay index at different times are shown in Figure \ref{fig:decay index}. The 
blue, green and white contours represent the values of decay index at 1.1, 1.5 and 2, 
respectively. Panels (d) and (e) show two different 3D views at 17:46:25 
UT, where the orange plane represents the distribution of the decay index on the X-Y 
plane above the highest flux rope. It can be seen that the flux ropes are very close to 
the contour of the decay index 1.5. From the positions of the top of the flux rope 
closest to the $n = 1.5$ contour at different times, we show the X-Z plane that cuts the 
X-Y plane at S1 (Figure \ref{fig:decay index}(e)) for the time 17:46:25 UT and 17:58:25 
UT, and the X-Z plane at S2 for 18:10:25 UT. The distribution of the decay index of the 
three different planes and the position of the flux ropes are shown in Figures 
\ref{fig:decay index}(a)-(c). The highest pink flux ropes do not intersect with the 
contour of $n = 1.5$ until 18:10:25 UT, so the extrapolation results suggest that the 
seed hot channels do not reach the threshold of torus instability before the impulsive 
acceleration. Combined with the fact that the onset of the seed hot channels 
acceleration starts earlier than the associated flare, the fast ``flare reconnection" 
unlikely triggers the hot channels acceleration \citep{Cheng2020}. In summary, the 
reconnection during the second flare precursor perhaps contributes to the 
initiation of the hot channels impulsive rise.

\subsection{Driving Mechanisms of the Observed Writhing Motion}

 When the twist of the MFR exceeds the critical value (approximately 3.5 $\pi$, 
 i.e. 1.75 turn, but changes with different aspect ratio of the loops involved), the
 kink instability will occur and part of the twist of the MFR will be transformed into 
 writhe \citep{Baty2001,Torok2014}. In a data-constrained MHD simulation, 
\citet{Inoue2018} have found that a series of small flux ropes reconnect with each 
other in the early stage of the  eruption, forming a large and highly twisted flux rope, 
which is similar to the first kinking hot channel in our event. Both hot channels in 
this event show a writhing motion during the eruption. We calculate the 
photospheric magnetic flux of the area surrounded by the green box in 
Figure \ref{fig:flux rope}(d), which is large enough to include the whole left 
footpoint area of the flux rope. The temporal evolution of magnetic flux in the region 
with $T_{w} \le -1.0$ and $T_{w} \le -1.75$, and the evolution of the ratio of magnetic 
flux in the region with $T_{w} \le -1.75$ to that in the region with $T_{w} \le -1.0$ 
are shown in Figure \ref{fig:decay index}(f). The results show that the magnetic flux in 
the region with $T_{w} \le -1.75$ increases rapidly after 17:34:25 UT, while the 
magnetic flux in the region with $T_{w} \le -1.0$ decreases 
after 17:34:25 UT. The ratio of the magnetic flux in the region with $T_{w} \le -1.75$ 
to the magnetic flux in the region with $T_{w} \le -1.0$ increases obviously, but the 
proportion in the region with $T_{w} \le -1.75$ does not exceed 30$\%$ both before and 
after the writhing motion of the flux rope. Therefore, KI may not be the main driver of 
writhing motion of the first kinking hot channel in this event \citep{Inoue2018}. 

Once the flux rope starts to rise up, it is out of its initial equilibrium, and the 
Lorentz force becomes non-zero and can act on the MFR and rotate it \citep{Isenberg2007}.
This can be reflected by the increase of the total current in the cross section of the 
flux rope during its rising \citep{Inoue2018}. Figure \ref{fig:decay index}(g) shows 
that the evolutions of total current vertically crossing the footpoint of the MFR
(marked by the green box in Figure \ref{fig:flux rope}(d)) in the region with 
$T_{w} \le -1.75$ and $T_{w} \le -1$, which are similar to the evolution of magnetic 
flux. The total current in the region with $T_{w} \le -1.75$ begins to increase rapidly 
at 17:34 UT and decreases rapidly after the writhing motion of the hot channel. This 
suggests that the Lorentz force might contribute to the writhing motion of MFR. In 
addition, during the rise of the flux rope, the magnetic tension of the twisted magnetic 
fields will be released, and whether KI is triggered or not, the release of magnetic 
tension can also contribute to the writhing of the flux rope axis \citep{Kliem2012}. 
Therefore, the writhing motion of the first kinking hot channel may be driven by a combination of these two mechanisms.

\section{Summary and Discussions}

We investigate the formation and eruption of hot channels during the M6.5 class 
flare occurring on 2015 June 22 using both observations and NLFFF extrapolations. 
There are two precursors before the flare main phase. The seed hot channels appear
after the onset of the first precursor and grow gradually. During the second precursor,
the footpoints' position of the seed hot channels hardly changes, but the width of the
seed hot channels increases significantly. After the peak of the second 
precursor, the impulsive acceleration of the hot channels begins. In the process of 
acceleration, merge reconnection between different seed hot channels likely occurs, 
forming a longer and more twisted flux rope. The newly formed hot channel soon evolves 
into an obvious kinking structure. After the first kinking hot channel disappears, the 
second hot channel appears with an existing kinking structure at an adjacent location. 
The eruption of these two hot channels produce two peaks on the flare's GOES light 
curve, which is similar to \citet{Wang2018b}. In this study we focus on the 
formation and writhing motion of the first hot channel, and those of the second hot 
channel is unclear due to the strong background emission.

With the appearance and buildup of the hot channels, footpoint brightenings motion 
parallel to and perpendicular to the PIL are observed by SDO/AIA. \citet{Priest2017} 
proposed that magnetic reconnection has two phases: the 3D ``zipper reconnection"  
between sheared arcades related to the elongation of the flare ribbons along the PIL, 
and the quasi-2D ``main phase reconnection" of unsheared fields around the flux rope 
related to the expansion of flare ribbons away from the PIL. In the current event, the 
motion of brightenings along the PIL is observed during two flare precursors and the 
beginning of the flare main phase, and the expansion of brightening perpendicular to the 
PIL is observed after the onset of the flare main phase. During the first precursor of 
this event, the brightenings on the two sides of the PIL mainly show parallel 
motion toward the north, which may correspond to the ``simple zippettes" reconnection 
between sheared arcades, and leads to the formation of flux ropes \citep{Priest2017}. 
The brightenings on the two sides of the PIL mainly show parallel motion toward the 
north during the second precursor, and this observation may be due to the occurrence of 
``helical zippettes" reconnection on the foundation of the flux rope formed during the 
first precursor, which results in the further twist accumulation in the flux ropes. 
From the late stage of the second flare precursor to the impulsive rise of the flare, 
the observed simultaneous southward and northward converging motion of footpoint 
brightenings may correspond to the occurrence of ``converging reconnection" between the 
flux ropes formed during the two flare precursors. That is, these flux ropes reconnect 
into a new flux rope which is much longer and has a much stronger twist
\citep{Priest2017}. The converging reconnection is similar to the ``rr-fr" reconnection 
that involves two flux-rope field lines that reconnect in to another multi-turn 
flux-rope field line and a flare loop proposed by \citet{Aulanier2019}. In 
addition to the footpoint motion, the three jets observed during the two flare 
precursors and flare main phase accompanying the footpoint motion also support the 
occurrence of magnetic reconnection.

We have constructed a series of NLFFF extrapolations based on the single, 
isolated vector magnetograms observed by SDO/HMI before and during the eruption. The 
extrapolations incorporate no information about the prior evolution of the photospheric 
and coronal magnetic field \citep{Cheung2012}. One important condition for using NLFFF 
is that the magnetic field must evolve slowly compared to the Alfv'en crossing time 
\citep{Savcheva2012}. Therefore, our NLFFF extrapolations can model the buildup phase of 
a flux rope before the eruption, but they fail in correctly describing the intrinsic 
dynamic evolution of the magnetic configuration and plasma properties, since these 
magnetic fields are independent of each other \citep{Kliem2013,Jiang2014}. Our NLFFF 
extrapolations show that the mean and maximum values of $T_{w}$ in the region with 
$T_{w} \le -1.75$ increase significantly during both flare precursors and main 
phases, and reach a maximum after the onset of the flare main phase. These results 
suggest the continuous twist accumulation through difference phases, which is 
consistent with the increase of twist of the flux rope in the envisaged zipper 
reconnection process.  A series of NLFFF extrapolations show that two flux ropes
with negative twist gradually widen and hence approach each other and only one 
flux rope is identified after the eruption onset. In the mean time, the AIA 131~\AA 
observations show that seed hot channels gradual approach and merge into a much longer 
and more twisted hot channel, which is consistent with the theory that parallel flux 
tube with the same twist can merge into a single flux tube \citep{Linton2001}. Thus both 
observations and extrapolations imply the occurrence of zipper reconnection and 
merging between seed hot channels.

The impulsive acceleration of the hot channels begins after the peak of the second flare 
precursor, which suggests that magnetic reconnection may contribute to the initiation of 
the impulsive acceleration of the hot channels. The relative position between the 
selected flux rope filed lines and the decay index distribution on the X-Z section 
indicates the height of flux rope does not reach the critical height of TI until about 
20 minutes after the beginning of impulsive acceleration. Therefore, the onset of 
impulsive acceleration of the hot channels is unlikely driven by TI, which is consistent 
with the conclusion of \citet{Kang2019}. Their study also suggests that the 
system is unstable against the DAI caused by the additional upward Lorentz force on the 
bent of double-arc current loop, and it can happen even if the system does not reach the 
critical of TI \citep{Ishiguro2017,Kusano2020}. However, our study suggests that 
magnetic reconnection during the flare precursors may also play an important role in the 
onset of the impulsive acceleration of the hot channels.

Our results can be summarized as follows: (1) The seed hot channels appear and build up
during the two flare precursors, and at these stages the footpoint motion parallel to 
the PIL suggests the occurrence of zipper reconnection. (2) ``Simple zippettes" 
reconnection during the first flare precursor leads to the formation of the seed hot 
channels and contributes to the continuous buildup. ``Helical zippettes" 
reconnection during the second flare precursor plays an important role in the onset of 
the impulsive acceleration of the hot channels. (3) The merging between the seed hot 
channels leads to the substantial twist buildup of the first kinking hot channel, which 
shows writhing motion during its fast rise. The increase of Lorentz force is identified 
associated with the writhing motion, which may be driven by the combined effect of the 
Lorentz force induced by the external sheared field and the magnetic tension release of 
the twisted field. (4) Unlikely driven by TI, the impulsive acceleration of the hot 
channels may be attributed to magnetic reconnection during the second flare precursor.

\acknowledgments
The authors thank the referee for providing constructive suggestions to improve the 
paper. We also thank the SDO teams for providing the valuable data. This work is 
supported by the National Key R\&D Program of China 2021YFA1600500 
(2021YFA1600502), the Chinese foundations NSFC (12173092, 41761134088, 11790302 
(11790300), U1731241, 41774150, 11925302, and 42188101), and the Strategic Priority 
Research Program onSpace Science, CAS, Grant No. XDA15052200 and XDA15320301.

\bibliography{paper2}{}

\begin{thebibliography}{}
\expandafter\ifx\csname natexlab\endcsname\relax\def\natexlab#1{#1}\fi
\providecommand{\url}[1]{\href{#1}{#1}}
\providecommand{\dodoi}[1]{doi:~\href{http://doi.org/#1}{\nolinkurl{#1}}}
\providecommand{\doeprint}[1]{\href{http://ascl.net/#1}{\nolinkurl{http://ascl.net/#1}}}
\providecommand{\doarXiv}[1]{\href{https://arxiv.org/abs/#1}{\nolinkurl{https://arxiv.org/abs/#1}}}

\bibitem[{{Antiochos} {et~al.}(1999){Antiochos}, {DeVore}, \&
  {Klimchuk}}]{Antiochos1999}
{Antiochos}, S.~K., {DeVore}, C.~R., \& {Klimchuk}, J.~A. 1999, \apj, 510, 485,
  \dodoi{10.1086/306563}

\bibitem[{{Aulanier} \& {Dud{\'\i}k}(2019)}]{Aulanier2019}
{Aulanier}, G., \& {Dud{\'\i}k}, J. 2019, \aap, 621, A72,
  \dodoi{10.1051/0004-6361/201834221}

\bibitem[{{Aulanier} {et~al.}(2012){Aulanier}, {Janvier}, \&
  {Schmieder}}]{Aulanier2012}
{Aulanier}, G., {Janvier}, M., \& {Schmieder}, B. 2012, \aap, 543, A110,
  \dodoi{10.1051/0004-6361/201219311}

\bibitem[{{Aulanier} {et~al.}(2010){Aulanier}, {T{\"o}r{\"o}k}, {D{\'e}moulin},
  \& {DeLuca}}]{Aulanier2010}
{Aulanier}, G., {T{\"o}r{\"o}k}, T., {D{\'e}moulin}, P., \& {DeLuca}, E.~E.
  2010, \apj, 708, 314, \dodoi{10.1088/0004-637X/708/1/314}

\bibitem[{{Awasthi} {et~al.}(2018){Awasthi}, {Liu}, {Wang}, {Wang}, \&
  {Shen}}]{Awasthi2018}
{Awasthi}, A.~K., {Liu}, R., {Wang}, H., {Wang}, Y., \& {Shen}, C. 2018, \apj,
  857, 124, \dodoi{10.3847/1538-4357/aab7fb}

\bibitem[{{Baty}(2001)}]{Baty2001}
{Baty}, H. 2001, \aap, 367, 321, \dodoi{10.1051/0004-6361:20000412}

\bibitem[{{Bi} {et~al.}(2017){Bi}, {Yang}, {Jiang}, {Hong}, {Xu}, {Qu}, \&
  {Ji}}]{Bi2017}
{Bi}, Y., {Yang}, J., {Jiang}, Y., {et~al.} 2017, \apjl, 849, L35,
  \dodoi{10.3847/2041-8213/aa960e}

\bibitem[{{Cao} {et~al.}(2010){Cao}, {Gorceix}, {Coulter}, {Ahn}, {Rimmele}, \&
  {Goode}}]{Cao2010}
{Cao}, W., {Gorceix}, N., {Coulter}, R., {et~al.} 2010, Astronomische
  Nachrichten, 331, 636, \dodoi{10.1002/asna.201011390}

\bibitem[{{Cheng} \& {Ding}(2016)}]{Cheng2016}
{Cheng}, X., \& {Ding}, M.~D. 2016, \apjs, 225, 16,
  \dodoi{10.3847/0067-0049/225/1/16}

\bibitem[{{Cheng} {et~al.}(2017){Cheng}, {Guo}, \& {Ding}}]{Cheng2017}
{Cheng}, X., {Guo}, Y., \& {Ding}, M. 2017, Science China Earth Sciences, 60,
  1383, \dodoi{10.1007/s11430-017-9074-6}

\bibitem[{{Cheng} {et~al.}(2013{\natexlab{a}}){Cheng}, {Zhang}, {Ding}, {Liu},
  \& {Poomvises}}]{Cheng2013b}
{Cheng}, X., {Zhang}, J., {Ding}, M.~D., {Liu}, Y., \& {Poomvises}, W.
  2013{\natexlab{a}}, \apj, 763, 43, \dodoi{10.1088/0004-637X/763/1/43}

\bibitem[{{Cheng} {et~al.}(2013{\natexlab{b}}){Cheng}, {Zhang}, {Ding},
  {Olmedo}, {Sun}, {Guo}, \& {Liu}}]{Cheng2013a}
{Cheng}, X., {Zhang}, J., {Ding}, M.~D., {et~al.} 2013{\natexlab{b}}, \apjl,
  769, L25, \dodoi{10.1088/2041-8205/769/2/L25}

\bibitem[{{Cheng} {et~al.}(2020){Cheng}, {Zhang}, {Kliem}, {T{\"o}r{\"o}k},
  {Xing}, {Zhou}, {Inhester}, \& {Ding}}]{Cheng2020}
{Cheng}, X., {Zhang}, J., {Kliem}, B., {et~al.} 2020, \apj, 894, 85,
  \dodoi{10.3847/1538-4357/ab886a}

\bibitem[{{Cheng} {et~al.}(2011){Cheng}, {Zhang}, {Liu}, \& {Ding}}]{Cheng2011}
{Cheng}, X., {Zhang}, J., {Liu}, Y., \& {Ding}, M.~D. 2011, \apjl, 732, L25,
  \dodoi{10.1088/2041-8205/732/2/L25}

\bibitem[{{Cheung} \& {DeRosa}(2012)}]{Cheung2012}
{Cheung}, M. C.~M., \& {DeRosa}, M.~L. 2012, \apj, 757, 147,
  \dodoi{10.1088/0004-637X/757/2/147}

\bibitem[{{Fan}(2009)}]{Fan2009}
{Fan}, Y. 2009, \apj, 697, 1529, \dodoi{10.1088/0004-637X/697/2/1529}

\bibitem[{{Fan} \& {Gibson}(2003)}]{Fan2003}
{Fan}, Y., \& {Gibson}, S.~E. 2003, \apjl, 589, L105, \dodoi{10.1086/375834}

\bibitem[{{Fan} \& {Gibson}(2004)}]{Fan2004}
---. 2004, \apj, 609, 1123, \dodoi{10.1086/421238}

\bibitem[{{Fan} \& {Gibson}(2007)}]{Fan2007}
---. 2007, \apj, 668, 1232, \dodoi{10.1086/521335}

\bibitem[{{Forbes} \& {Priest}(1995)}]{Forbes1995}
{Forbes}, T.~G., \& {Priest}, E.~R. 1995, \apj, 446, 377,
  \dodoi{10.1086/175797}

\bibitem[{{Gou} {et~al.}(2019){Gou}, {Liu}, {Kliem}, {Wang}, \&
  {Veronig}}]{Gou2019}
{Gou}, T., {Liu}, R., {Kliem}, B., {Wang}, Y., \& {Veronig}, A.~M. 2019,
  Science Advances, 5, 7004, \dodoi{10.1126/sciadv.aau7004}

\bibitem[{{Green} \& {Kliem}(2009)}]{Green2009}
{Green}, L.~M., \& {Kliem}, B. 2009, \apjl, 700, L83,
  \dodoi{10.1088/0004-637X/700/2/L83}

\bibitem[{{Green} \& {Kliem}(2014)}]{Green&Kliem2014}
{Green}, L.~M., \& {Kliem}, B. 2014, in Nature of Prominences and their Role in
  Space Weather, ed. B.~{Schmieder}, J.-M. {Malherbe}, \& S.~T. {Wu}, Vol. 300,
  209--214, \dodoi{10.1017/S1743921313010983}

\bibitem[{{Green} {et~al.}(2011){Green}, {Kliem}, \& {Wallace}}]{Green2011}
{Green}, L.~M., {Kliem}, B., \& {Wallace}, A.~J. 2011, \aap, 526, A2,
  \dodoi{10.1051/0004-6361/201015146}

\bibitem[{{Inoue} {et~al.}(2018){Inoue}, {Shiota}, {Bamba}, \&
  {Park}}]{Inoue2018}
{Inoue}, S., {Shiota}, D., {Bamba}, Y., \& {Park}, S.-H. 2018, \apj, 867, 83,
  \dodoi{10.3847/1538-4357/aae079}

\bibitem[{{Isenberg} \& {Forbes}(2007)}]{Isenberg2007}
{Isenberg}, P.~A., \& {Forbes}, T.~G. 2007, \apj, 670, 1453,
  \dodoi{10.1086/522025}

\bibitem[{{Ishiguro} \& {Kusano}(2017)}]{Ishiguro2017}
{Ishiguro}, N., \& {Kusano}, K. 2017, \apj, 843, 101,
  \dodoi{10.3847/1538-4357/aa799b}

\bibitem[{{Jiang} {et~al.}(2014){Jiang}, {Wu}, {Feng}, \& {Hu}}]{Jiang2014}
{Jiang}, C., {Wu}, S.~T., {Feng}, X., \& {Hu}, Q. 2014, \apj, 780, 55,
  \dodoi{10.1088/0004-637X/780/1/55}

\bibitem[{{Jiang} {et~al.}(2021){Jiang}, {Feng}, {Liu}, {Yan}, {Hu}, {Moore},
  {Duan}, {Cui}, {Zuo}, {Wang}, \& {Wei}}]{Jiang2021}
{Jiang}, C., {Feng}, X., {Liu}, R., {et~al.} 2021, Nature Astronomy, 5, 1126,
  \dodoi{10.1038/s41550-021-01414-z}

\bibitem[{{Jing} {et~al.}(2017){Jing}, {Liu}, {Cheung}, {Lee}, {Xu}, {Liu},
  {Zhu}, \& {Wang}}]{Jing2017}
{Jing}, J., {Liu}, R., {Cheung}, M. C.~M., {et~al.} 2017, \apjl, 842, L18,
  \dodoi{10.3847/2041-8213/aa774d}

\bibitem[{{Jing} {et~al.}(2016){Jing}, {Xu}, {Cao}, {Liu}, {Gary}, \&
  {Wang}}]{Jing2016}
{Jing}, J., {Xu}, Y., {Cao}, W., {et~al.} 2016, Scientific Reports, 6, 24319,
  \dodoi{10.1038/srep24319}

\bibitem[{{Kang} {et~al.}(2019){Kang}, {Inoue}, {Kusano}, {Park}, \&
  {Moon}}]{Kang2019}
{Kang}, J., {Inoue}, S., {Kusano}, K., {Park}, S.-H., \& {Moon}, Y.-J. 2019,
  \apj, 887, 263, \dodoi{10.3847/1538-4357/ab5582}

\bibitem[{{Karpen} {et~al.}(2012){Karpen}, {Antiochos}, \&
  {DeVore}}]{Karpen2012}
{Karpen}, J.~T., {Antiochos}, S.~K., \& {DeVore}, C.~R. 2012, \apj, 760, 81,
  \dodoi{10.1088/0004-637X/760/1/81}

\bibitem[{{Kliem} {et~al.}(2013){Kliem}, {Su}, {van Ballegooijen}, \&
  {DeLuca}}]{Kliem2013}
{Kliem}, B., {Su}, Y.~N., {van Ballegooijen}, A.~A., \& {DeLuca}, E.~E. 2013,
  \apj, 779, 129, \dodoi{10.1088/0004-637X/779/2/129}

\bibitem[{{Kliem} \& {T{\"o}r{\"o}k}(2006)}]{Kliem&Torok2006}
{Kliem}, B., \& {T{\"o}r{\"o}k}, T. 2006, \prl, 96, 255002,
  \dodoi{10.1103/PhysRevLett.96.255002}

\bibitem[{{Kliem} {et~al.}(2012){Kliem}, {T{\"o}r{\"o}k}, \&
  {Thompson}}]{Kliem2012}
{Kliem}, B., {T{\"o}r{\"o}k}, T., \& {Thompson}, W.~T. 2012, \solphys, 281,
  137, \dodoi{10.1007/s11207-012-9990-z}

\bibitem[{{Kopp} \& {Pneuman}(1976)}]{Kopp1976}
{Kopp}, R.~A., \& {Pneuman}, G.~W. 1976, \solphys, 50, 85,
  \dodoi{10.1007/BF00206193}

\bibitem[{{Krucker} {et~al.}(2005){Krucker}, {Fivian}, \& {Lin}}]{Krucker2005}
{Krucker}, S., {Fivian}, M.~D., \& {Lin}, R.~P. 2005, Advances in Space
  Research, 35, 1707, \dodoi{10.1016/j.asr.2005.05.054}

\bibitem[{{Kusano} {et~al.}(2020){Kusano}, {Iju}, {Bamba}, \&
  {Inoue}}]{Kusano2020}
{Kusano}, K., {Iju}, T., {Bamba}, Y., \& {Inoue}, S. 2020, Science, 369, 587,
  \dodoi{10.1126/science.aaz2511}

\bibitem[{{Lemen} {et~al.}(2012){Lemen}, {Title}, {Akin}, {Boerner}, {Chou},
  {Drake}, {Duncan}, {Edwards}, {Friedlaender}, {Heyman}, {Hurlburt}, {Katz},
  {Kushner}, {Levay}, {Lindgren}, {Mathur}, {McFeaters}, {Mitchell}, {Rehse},
  {Schrijver}, {Springer}, {Stern}, {Tarbell}, {Wuelser}, {Wolfson}, {Yanari},
  {Bookbinder}, {Cheimets}, {Caldwell}, {Deluca}, {Gates}, {Golub}, {Park},
  {Podgorski}, {Bush}, {Scherrer}, {Gummin}, {Smith}, {Auker}, {Jerram},
  {Pool}, {Soufli}, {Windt}, {Beardsley}, {Clapp}, {Lang}, \&
  {Waltham}}]{Lemen2012}
{Lemen}, J.~R., {Title}, A.~M., {Akin}, D.~J., {et~al.} 2012, \solphys, 275,
  17, \dodoi{10.1007/s11207-011-9776-8}

\bibitem[{{Li} \& {Zhang}(2009)}]{Li2009}
{Li}, L., \& {Zhang}, J. 2009, \apj, 690, 347,
  \dodoi{10.1088/0004-637X/690/1/347}

\bibitem[{{Linton} {et~al.}(2001){Linton}, {Dahlburg}, \&
  {Antiochos}}]{Linton2001}
{Linton}, M.~G., {Dahlburg}, R.~B., \& {Antiochos}, S.~K. 2001, \apj, 553, 905,
  \dodoi{10.1086/320974}

\bibitem[{{Liu} {et~al.}(2006){Liu}, {Lee}, {Deng}, {Gary}, \&
  {Wang}}]{Liu2006}
{Liu}, C., {Lee}, J., {Deng}, N., {Gary}, D.~E., \& {Wang}, H. 2006, \apj, 642,
  1205, \dodoi{10.1086/501000}

\bibitem[{{Liu} {et~al.}(2016{\natexlab{a}}){Liu}, {Xu}, {Cao}, {Deng}, {Lee},
  {Hudson}, {Gary}, {Wang}, {Jing}, \& {Wang}}]{Liu2016a}
{Liu}, C., {Xu}, Y., {Cao}, W., {et~al.} 2016{\natexlab{a}}, Nature
  Communications, 7, 13104, \dodoi{10.1038/ncomms13104}

\bibitem[{{Liu} {et~al.}(2018){Liu}, {Cheng}, {Wang}, {Zhou}, {Guo}, \&
  {Cui}}]{Liu2018a}
{Liu}, L., {Cheng}, X., {Wang}, Y., {et~al.} 2018, \apjl, 867, L5,
  \dodoi{10.3847/2041-8213/aae826}

\bibitem[{{Liu}(2020)}]{Liu2020}
{Liu}, R. 2020, Research in Astronomy and Astrophysics, 20, 165,
  \dodoi{10.1088/1674-4527/20/10/165}

\bibitem[{{Liu} {et~al.}(2010){Liu}, {Liu}, {Wang}, {Deng}, \&
  {Wang}}]{Liu2010}
{Liu}, R., {Liu}, C., {Wang}, S., {Deng}, N., \& {Wang}, H. 2010, \apjl, 725,
  L84, \dodoi{10.1088/2041-8205/725/1/L84}

\bibitem[{{Liu} {et~al.}(2016{\natexlab{b}}){Liu}, {Kliem}, {Titov}, {Chen},
  {Wang}, {Wang}, {Liu}, {Xu}, \& {Wiegelmann}}]{Liu2016b}
{Liu}, R., {Kliem}, B., {Titov}, V.~S., {et~al.} 2016{\natexlab{b}}, \apj, 818,
  148, \dodoi{10.3847/0004-637X/818/2/148}

\bibitem[{{Liu} {et~al.}(2022){Liu}, {Su}, {Liu}, {Chen}, {Liu}, \&
  {Ji}}]{Liu2022}
{Liu}, Y., {Su}, Y., {Liu}, R., {et~al.} 2022, \apj, 930, 130,
  \dodoi{10.3847/1538-4357/ac63ac}

\bibitem[{{Low} \& {Hundhausen}(1995)}]{Low1995}
{Low}, B.~C., \& {Hundhausen}, J.~R. 1995, \apj, 443, 818,
  \dodoi{10.1086/175572}

\bibitem[{{Lynch} {et~al.}(2008){Lynch}, {Antiochos}, {DeVore}, {Luhmann}, \&
  {Zurbuchen}}]{Lynch2008}
{Lynch}, B.~J., {Antiochos}, S.~K., {DeVore}, C.~R., {Luhmann}, J.~G., \&
  {Zurbuchen}, T.~H. 2008, \apj, 683, 1192, \dodoi{10.1086/589738}

\bibitem[{{Lynch} {et~al.}(2009){Lynch}, {Antiochos}, {Li}, {Luhmann}, \&
  {DeVore}}]{Lynch2009}
{Lynch}, B.~J., {Antiochos}, S.~K., {Li}, Y., {Luhmann}, J.~G., \& {DeVore},
  C.~R. 2009, \apj, 697, 1918, \dodoi{10.1088/0004-637X/697/2/1918}

\bibitem[{{Mackay} {et~al.}(2010){Mackay}, {Karpen}, {Ballester}, {Schmieder},
  \& {Aulanier}}]{Mackay2010}
{Mackay}, D.~H., {Karpen}, J.~T., {Ballester}, J.~L., {Schmieder}, B., \&
  {Aulanier}, G. 2010, \ssr, 151, 333, \dodoi{10.1007/s11214-010-9628-0}

\bibitem[{{Moore} {et~al.}(2001){Moore}, {Sterling}, {Hudson}, \&
  {Lemen}}]{Moore2001}
{Moore}, R.~L., {Sterling}, A.~C., {Hudson}, H.~S., \& {Lemen}, J.~R. 2001,
  \apj, 552, 833, \dodoi{10.1086/320559}

\bibitem[{{O'Dwyer} {et~al.}(2010){O'Dwyer}, {Del Zanna}, {Mason}, {Weber}, \&
  {Tripathi}}]{O'Dwyer2010}
{O'Dwyer}, B., {Del Zanna}, G., {Mason}, H.~E., {Weber}, M.~A., \& {Tripathi},
  D. 2010, \aap, 521, A21, \dodoi{10.1051/0004-6361/201014872}

\bibitem[{{Panasenco} {et~al.}(2011){Panasenco}, {Martin}, {Joshi}, \&
  {Srivastava}}]{Panasenco2011}
{Panasenco}, O., {Martin}, S., {Joshi}, A.~D., \& {Srivastava}, N. 2011,
  Journal of Atmospheric and Solar-Terrestrial Physics, 73, 1129,
  \dodoi{10.1016/j.jastp.2010.09.010}

\bibitem[{{Patsourakos} {et~al.}(2013){Patsourakos}, {Vourlidas}, \&
  {Stenborg}}]{Patsourakos2013}
{Patsourakos}, S., {Vourlidas}, A., \& {Stenborg}, G. 2013, \apj, 764, 125,
  \dodoi{10.1088/0004-637X/764/2/125}

\bibitem[{{Patsourakos} {et~al.}(2020){Patsourakos}, {Vourlidas},
  {T{\"o}r{\"o}k}, {Kliem}, {Antiochos}, {Archontis}, {Aulanier}, {Cheng},
  {Chintzoglou}, {Georgoulis}, {Green}, {Leake}, {Moore}, {Nindos}, {Syntelis},
  {Yardley}, {Yurchyshyn}, \& {Zhang}}]{Patsourakos2020}
{Patsourakos}, S., {Vourlidas}, A., {T{\"o}r{\"o}k}, T., {et~al.} 2020, \ssr,
  216, 131, \dodoi{10.1007/s11214-020-00757-9}

\bibitem[{{Pesnell} {et~al.}(2012){Pesnell}, {Thompson}, \&
  {Chamberlin}}]{Pesnell2012}
{Pesnell}, W.~D., {Thompson}, B.~J., \& {Chamberlin}, P.~C. 2012, \solphys,
  275, 3, \dodoi{10.1007/s11207-011-9841-3}

\bibitem[{{Priest} \& {Longcope}(2017)}]{Priest2017}
{Priest}, E.~R., \& {Longcope}, D.~W. 2017, \solphys, 292, 25,
  \dodoi{10.1007/s11207-016-1049-0}

\bibitem[{{Qiu}(2009)}]{Qiu2009}
{Qiu}, J. 2009, \apj, 692, 1110, \dodoi{10.1088/0004-637X/692/2/1110}

\bibitem[{{Rust} \& {Kumar}(1994)}]{Rust1994}
{Rust}, D.~M., \& {Kumar}, A. 1994, \solphys, 155, 69,
  \dodoi{10.1007/BF00670732}

\bibitem[{{Sahu} {et~al.}(2020){Sahu}, {Joshi}, {Mitra}, {Veronig}, \&
  {Yurchyshyn}}]{Sahu2020}
{Sahu}, S., {Joshi}, B., {Mitra}, P.~K., {Veronig}, A.~M., \& {Yurchyshyn}, V.
  2020, \apj, 897, 157, \dodoi{10.3847/1538-4357/ab962b}

\bibitem[{{Savcheva} {et~al.}(2012){Savcheva}, {van Ballegooijen}, \&
  {DeLuca}}]{Savcheva2012}
{Savcheva}, A.~S., {van Ballegooijen}, A.~A., \& {DeLuca}, E.~E. 2012, \apj,
  744, 78, \dodoi{10.1088/0004-637X/744/1/78}

\bibitem[{{Schou} {et~al.}(2012){Schou}, {Scherrer}, {Bush}, {Wachter},
  {Couvidat}, {Rabello-Soares}, {Bogart}, {Hoeksema}, {Liu}, {Duvall}, {Akin},
  {Allard}, {Miles}, {Rairden}, {Shine}, {Tarbell}, {Title}, {Wolfson},
  {Elmore}, {Norton}, \& {Tomczyk}}]{Schou2012}
{Schou}, J., {Scherrer}, P.~H., {Bush}, R.~I., {et~al.} 2012, \solphys, 275,
  229, \dodoi{10.1007/s11207-011-9842-2}

\bibitem[{{Shibata} {et~al.}(1995){Shibata}, {Masuda}, {Shimojo}, {Hara},
  {Yokoyama}, {Tsuneta}, {Kosugi}, \& {Ogawara}}]{Shibata1995}
{Shibata}, K., {Masuda}, S., {Shimojo}, M., {et~al.} 1995, \apjl, 451, L83,
  \dodoi{10.1086/309688}

\bibitem[{{Shiota} {et~al.}(2010){Shiota}, {Kusano}, {Miyoshi}, \&
  {Shibata}}]{Shiota2010}
{Shiota}, D., {Kusano}, K., {Miyoshi}, T., \& {Shibata}, K. 2010, \apj, 718,
  1305, \dodoi{10.1088/0004-637X/718/2/1305}

\bibitem[{{Su} {et~al.}(2007){Su}, {Golub}, \& {Van Ballegooijen}}]{Su2007}
{Su}, Y., {Golub}, L., \& {Van Ballegooijen}, A.~A. 2007, \apj, 655, 606,
  \dodoi{10.1086/510065}

\bibitem[{{Su} \& {van Ballegooijen}(2012)}]{Su2012}
{Su}, Y., \& {van Ballegooijen}, A. 2012, \apj, 757, 168,
  \dodoi{10.1088/0004-637X/757/2/168}

\bibitem[{{Su} {et~al.}(2006){Su}, {Golub}, {van Ballegooijen}, \&
  {Gros}}]{Su2006}
{Su}, Y.~N., {Golub}, L., {van Ballegooijen}, A.~A., \& {Gros}, M. 2006,
  \solphys, 236, 325, \dodoi{10.1007/s11207-006-0039-z}

\bibitem[{{Titov} \& {D{\'e}moulin}(1999)}]{Titov1999}
{Titov}, V.~S., \& {D{\'e}moulin}, P. 1999, \aap, 351, 707

\bibitem[{{T{\"o}r{\"o}k} {et~al.}(2010){T{\"o}r{\"o}k}, {Berger}, \&
  {Kliem}}]{Torok2010}
{T{\"o}r{\"o}k}, T., {Berger}, M.~A., \& {Kliem}, B. 2010, \aap, 516, A49,
  \dodoi{10.1051/0004-6361/200913578}

\bibitem[{{T{\"o}r{\"o}k} \& {Kliem}(2005)}]{Torok&Kliem2005}
{T{\"o}r{\"o}k}, T., \& {Kliem}, B. 2005, \apjl, 630, L97,
  \dodoi{10.1086/462412}

\bibitem[{{T{\"o}r{\"o}k} {et~al.}(2014){T{\"o}r{\"o}k}, {Kliem}, {Berger},
  {Linton}, {D{\'e}moulin}, \& {van Driel-Gesztelyi}}]{Torok2014}
{T{\"o}r{\"o}k}, T., {Kliem}, B., {Berger}, M.~A., {et~al.} 2014, Plasma
  Physics and Controlled Fusion, 56, 064012,
  \dodoi{10.1088/0741-3335/56/6/064012}

\bibitem[{{T{\"o}r{\"o}k} {et~al.}(2004){T{\"o}r{\"o}k}, {Kliem}, \&
  {Titov}}]{Torok2004}
{T{\"o}r{\"o}k}, T., {Kliem}, B., \& {Titov}, V.~S. 2004, \aap, 413, L27,
  \dodoi{10.1051/0004-6361:20031691}

\bibitem[{{Tripathi} {et~al.}(2006){Tripathi}, {Isobe}, \&
  {Mason}}]{Tripathi2006}
{Tripathi}, D., {Isobe}, H., \& {Mason}, H.~E. 2006, \aap, 453, 1111,
  \dodoi{10.1051/0004-6361:20064993}

\bibitem[{{van Ballegooijen} {et~al.}(1998){van Ballegooijen}, {Cartledge}, \&
  {Priest}}]{vanBallegooijen1998}
{van Ballegooijen}, A.~A., {Cartledge}, N.~P., \& {Priest}, E.~R. 1998, \apj,
  501, 866, \dodoi{10.1086/305823}

\bibitem[{{van Ballegooijen} \& {Martens}(1989)}]{vanBallegooijen1989}
{van Ballegooijen}, A.~A., \& {Martens}, P.~C.~H. 1989, \apj, 343, 971,
  \dodoi{10.1086/167766}

\bibitem[{{Wang} {et~al.}(2017){Wang}, {Liu}, {Ahn}, {Xu}, {Jing}, {Deng},
  {Huang}, {Liu}, {Kusano}, {Fleishman}, {Gary}, \& {Cao}}]{Wang2017}
{Wang}, H., {Liu}, C., {Ahn}, K., {et~al.} 2017, Nature Astronomy, 1, 0085,
  \dodoi{10.1038/s41550-017-0085}

\bibitem[{{Wang} {et~al.}(2018{\natexlab{a}}){Wang}, {Liu}, {Deng}, \&
  {Wang}}]{Wang2018a}
{Wang}, J., {Liu}, C., {Deng}, N., \& {Wang}, H. 2018{\natexlab{a}}, \apj, 853,
  143, \dodoi{10.3847/1538-4357/aaa712}

\bibitem[{{Wang} {et~al.}(2018{\natexlab{b}}){Wang}, {Su}, {Shen}, {Yang},
  {Cao}, \& {Ji}}]{Wang2018b}
{Wang}, Y., {Su}, Y., {Shen}, J., {et~al.} 2018{\natexlab{b}}, \apj, 859, 148,
  \dodoi{10.3847/1538-4357/aac0f7}

\bibitem[{{Wiegelmann}(2004)}]{Wiegelmann2004}
{Wiegelmann}, T. 2004, \solphys, 219, 87,
  \dodoi{10.1023/B:SOLA.0000021799.39465.36}

\bibitem[{{Wiegelmann} {et~al.}(2006){Wiegelmann}, {Inhester}, {Kliem},
  {Valori}, \& {Neukirch}}]{Wiegelmann2006}
{Wiegelmann}, T., {Inhester}, B., {Kliem}, B., {Valori}, G., \& {Neukirch}, T.
  2006, \aap, 453, 737, \dodoi{10.1051/0004-6361:20054751}

\bibitem[{{Yang} {et~al.}(2009){Yang}, {Cheng}, {Krucker}, {Lin}, \&
  {Ip}}]{Yang2009}
{Yang}, Y.-H., {Cheng}, C.~Z., {Krucker}, S., {Lin}, R.~P., \& {Ip}, W.~H.
  2009, \apj, 693, 132, \dodoi{10.1088/0004-637X/693/1/132}

\bibitem[{{Yurchyshyn} {et~al.}(2009){Yurchyshyn}, {Abramenko}, \&
  {Tripathi}}]{Yurchyshyn2009}
{Yurchyshyn}, V., {Abramenko}, V., \& {Tripathi}, D. 2009, \apj, 705, 426,
  \dodoi{10.1088/0004-637X/705/1/426}

\bibitem[{{Zhang} {et~al.}(2012){Zhang}, {Cheng}, \& {Ding}}]{Zhang2012}
{Zhang}, J., {Cheng}, X., \& {Ding}, M.-D. 2012, Nature Communications, 3, 747,
  \dodoi{10.1038/ncomms1753}

\bibitem[{{Zhou} {et~al.}(2022){Zhou}, {Jiang}, {Liu}, {Wang}, {Liu}, \&
  {Cui}}]{Zhou2022}
{Zhou}, Z., {Jiang}, C., {Liu}, R., {et~al.} 2022, \apjl, 927, L14,
  \dodoi{10.3847/2041-8213/ac5740}

\end{thebibliography}
\bibliographystyle{aasjournal}

\newpage

\end{document}